\documentclass[pdftex,twocolumn,epjc3]{svjour3}
\smartqed
\RequirePackage{graphicx}
\RequirePackage{mathptmx}      
\RequirePackage{flushend}
\RequirePackage[numbers,sort&compress]{natbib}
\RequirePackage[colorlinks,citecolor=blue,urlcolor=blue,linkcolor=blue]{hyperref}

\DeclareSymbolFont{newfont}{OML}{cmm}{m}{it}
\DeclareMathSymbol{\Epsilon}{3}{newfont}{15}
\DeclareMathSymbol{\Varrho}{3}{newfont}{37}

\usepackage{graphicx,epsfig}
\usepackage{amsmath,mathrsfs,amsfonts}
\usepackage{amssymb}
\usepackage{longtable}
\usepackage{multirow}
\usepackage{enumerate}
\usepackage{dcolumn}
\usepackage{bm}
\usepackage{amsfonts}
\usepackage{subfigure}
\usepackage{color}

\newcommand{\be}{\begin{equation}}
\newcommand{\ee}{\end{equation}}
\newcommand{\bea}{\begin{eqnarray}}
\newcommand{\eea}{\end{eqnarray}}

\newcommand{\abs}[1]{\lvert#1\rvert}

\journalname{Eur. Phys. J. C}

\begin{document}

\title{Cosmological dynamics of dark energy in scalar-torsion $f(T,\phi)$ gravity}

\author{Manuel Gonzalez-Espinoza\thanksref{e1,addr1}
        \and
        Giovanni Otalora\thanksref{e2,addr1} 
}

\thankstext{e1}{e-mail: manuel.gonzalez@pucv.cl}
\thankstext{e2}{e-mail: giovanni.otalora@pucv.cl}

\institute{Instituto de F\'{\i}sica, Pontificia Universidad Cat\'olica de 
Valpara\'{\i}so, Casilla 4950, Valpara\'{\i}so, Chile\label{addr1}
}

\date{\today}

\maketitle

\begin{abstract}
It is investigated the cosmological dynamics of scalar-torsion $f(T,\phi)$ gravity as a dark energy model, where $T$ is the torsion scalar of teleparallel gravity and $\phi$ is a canonical scalar field. In this context, we are concerned with the phenomenology of the class of models with non-linear coupling to gravity and exponential potential. We obtain the critical points of the autonomous system, along with the stability conditions of each one of them and their cosmological properties. Particularly, we show the existence of new attractors with accelerated expansion, as well as, new scaling solutions in which the energy density of dark energy scales as the background fluid density, thus, defining the so-called scaling radiation and scaling matter epochs. The scaling solutions are saddle points, and therefore, the system exits these solutions to the current epoch of cosmic acceleration, towards an attractor point describing the dark energy-dominated era.  
\end{abstract}

\section{Introduction}\label{Introduction}

The discovery that the Universe is expanding at an accelerated rate, through the analysis of observational data of supernovas Ia \cite{Riess:1998cb,Perlmutter:1998np}, radically modified our understanding of Cosmology because it indicated the existence of a new component that constitutes $68\%$ of the total energy density our Universe. Even more, this new component remains a mystery due to the fact that its true nature is still unknown, and that is why it has been dubbed dark energy. Moreover, although the standard cosmology presents us with an excellent model when fitting the current observational data, by assuming the cosmological constant at the Einstein equations as the responsible for the accelerated expansion of our Universe, this assumption faces a severe fine tuning problem related with its energy scale, the so-called cosmological constant problem \cite{Weinberg:1988cp,Carroll:2000fy,Padilla:2015aaa}. In fact, the energy density associated with the cosmological constant today is to be of the order of the critical density, $\rho_{\Lambda}\sim 10^{-47} GeV^4$, but if we identify it with the energy density of the vacuum in quantum field theory, it should be enormously larger, about $10^{121}$ times larger than the observed value, that is, $\rho_{\Lambda} \sim 10^{74} GeV^4$ , when the cut-off scale is chosen to be the Planck scale \cite{Copeland:2006wr}. Furthermore, the latest observational data has pointed out some tensions or anomalies which are of statistical importance \cite{Aghanim:2018eyx,Sola:2017znb,Kazantzidis:2018rnb}. Particularly, the tension between the Planck experiment and other low-redshift probes at the measurement of the anisotropy of the Cosmic Microwave Background (CMB), the tension of the Hubble at the present time $H_0$ \cite{Riess:2011yx,Riess:2016jrr,Riess:2018byc,DiValentino:2020zio}, the tension at the measurement of the amplitude $\sigma_8$ and the growth rate of cosmic structure $f\sigma_8$ \cite{Hildebrandt:2016iqg,Kuijken:2015vca,Conti:2016gav,DiValentino:2018gcu,DiValentino:2020vvd}, etc.  Although this could mean systematic errors in the method to obtain the data, this also could indicate the necessity of a new cosmological model \cite{Riess:2019cxk,Davari:2019tni,DiValentino:2015bja,Sola:2019jek,Sola:2020lba,Joyce:2014kja,Koyama:2015vza}. 

As alternative theoretical constructions to address the cosmological constant problem we have at hand dynamical dark energy models with a modified matter source described by a scalar field such as quintessence \cite{Wetterich:1987fm,Ratra:1987rm,Carroll:1998zi,Tsujikawa:2013fta}, k-essence \cite{Chiba:1999ka,ArmendarizPicon:2000dh,ArmendarizPicon:2000ah}, Galileons \cite{Nicolis:2008in,Deffayet:2009wt,Baker:2017hug,Sakstein:2017xjx}, etc. The energy density of the scalar field evolves with time and, around the beginning of the radiation-dominated era, that value can be much larger than the observed value today for the energy density of dark energy, and then more compatible with energy scales of particle physics. Moreover, in theoretical physics the possibility of a non-minimal coupling to gravity cannot be excluded. This is motivated from quantum field theory in curved spacetimes where it can arise either through quantum corrections \cite{Linde:1982zj} or renormalizability requirements \cite{Freedman:1974gs,Freedman:1974ze,Birrell:1982ix}. In the context of scalar field cosmology, a non-minimal coupling to curvature scalar, it has been firstly studied in Ref. \cite{Perrotta:1999am}, and further investigated in Refs. \cite{Sahni:1998at,Chiba:1999wt,Bartolo:1999sq,Faraoni:2000wk,Hrycyna:2008gk,Hrycyna:2007gd}. For more developments in cosmology using non-minimally coupled scalar fields see for instance Ref. \cite{Copeland:2006wr,amendola2010dark}, and references therein. In scalar field cosmology, a very interesting and widely studied class of cosmological solutions are the scaling solutions.  For these solutions the energy density of the field decreases in proportion to the energy density of background fluid, and we can find them in scalar fields models with coupling to matter \cite{Amendola:1999er,Amendola:2006qi,Gomes:2013ema}, but also, when the non-minimal coupling to gravity is switched on \cite{Uzan:1999ch,Amendola:1999qq}. More interesting still, due to this special feature of the scaling solutions, the field energy density is not necessarily negligible compared to the energy density of the background fluid during early times, which allows to alleviate the aforementioned energy scale problem of the $\Lambda$CDM model \cite{Albuquerque:2018ymr,Ohashi:2009xw}. Finally, the minimally and non-minimally coupled scalar field models are giving good results for mitigating the current tensions in the concordance model \cite{DiValentino:2019ffd,DiValentino:2019jae}.

It is well known gravity can also be described in terms of torsion in the context of Teleparallel Gravity (TG) \cite{Einstein,TranslationEinstein,Early-papers1,Early-papers2,Early-papers3,Early-papers4,Early-papers5,Early-papers6,Aldrovandi-Pereira-book,JGPereira2,AndradeGuillenPereira-00,Arcos:2005ec,Pereira:2019woq}. In this theory, the dynamical variables are the tetrad field, instead of the usual metric tensor, and the Weitzenb\"{o}ck connection replaces the usual Levi-Civita connection \cite{Aldrovandi-Pereira-book,JGPereira2,AndradeGuillenPereira-00,Arcos:2005ec,Pereira:2019woq}. This produces a conceptual change as a result of using torsion instead of curvature, even though the field equations are equivalent, once the Lagrangian density of TG, the torsion scalar $T$, differs from the curvature scalar $R$ by a total derivative term. On the other hand, in the same way that we can propose a $f(R)$ gravity extension of GR \cite{Clifton:2011jh,Capozziello:2011et,DeFelice:2010aj,Nojiri:2010wj,Nojiri:2006ri}, we can promote the Lagrangian density of TG for a general function of the torsion scalar $T$ to obtain $f(T)$ gravity \cite{Bengochea:2008gz,Linder:2010py}. This latter belongs to a different class of modify gravity theories with distinctive features, for example, whereas $f(R)$ gravity is a fourth-order theory, $f(T)$ gravity has the advantage that its field equations are of second order. Additionally, since $f(T)$ gravity can explain the current accelerated expansion of our Universe, which has aroused great interest in these theories, it has also led to a fair number of investigations where it has been examined their several features, including observational solar system constraints \cite{Iorio:2012cm,Iorio:2015rla,Farrugia:2016xcw}, cosmological constraints \cite{Bengochea:2010sg, Wei:2011jw,Capozziello:2015rda,Oikonomou:2016jjh}, cosmological perturbations \cite{Dent:2011zz,Zheng:2010am,Izumi:2012qj,Li:2011wu,Peirone:2019yjs}, among others (for an extensive review see Ref. \cite{Cai:2015emx}). Even more, an important extension of $f(T)$ gravity is obtained by a non-minimal coupling between matter and torsion \cite{Harko:2014aja,Harko:2014sja,Carloni:2015lsa,Gonzalez-Espinoza:2018gyl}, when we consider an analogy with the non-minimal curvature-matter coupling in $f(R)$ gravity \cite{Nojiri:2004bi,Allemandi:2005qs,Nojiri:2006ri,Bertolami:2007gv,Harko:2008qz,Harko:2010mv,Bertolami:2009ic,Bertolami:2013kca,Wang:2013fja}, whose principal motivation are the counterterms that appear at the moment of quantising a scalar field with self-interaction at curved spacetime \cite{birrell1984quantum,Faraoni:1996rf,Faraoni:2000wk,Faraoni:2004pi}. Also, we can go one step further by considering a generalised teleparallel scalar-torsion $f(T,\phi)$ gravity \cite{Hohmann:2018rwf,Gonzalez-Espinoza:2020azh}, which for example encompasses $f(T)$ gravity with scalar field \cite{Yerzhanov:2010vu,Chakrabarti:2017moe,Rezazadeh:2015dza,Goodarzi:2018feh,Bamba:2016gbu}, non-minimally coupled scalar-torsion gravity \cite{Geng:2011aj,Geng:2011ka,Xu:2012jf,Wei:2011yr,Otalora:2013tba,Otalora:2013dsa,Otalora:2014aoa,Skugoreva:2014ena,Jarv:2015odu,Gonzalez-Espinoza:2019ajd}, and its extensions by including a non-linear scalar-torsion coupling \cite{Gonzalez-Espinoza:2020azh}. This latter kind of models are motivated by the already mentioned non-minimal torsion-matter coupling extension of $f(T)$ gravity \cite{Harko:2014aja,Harko:2014sja,Carloni:2015lsa,Gonzalez-Espinoza:2018gyl}, and its counterpart based on curvature, such as for example the case of a non-linear matter-curvature coupling in curvature-based modified gravity \cite{Nojiri:2004bi,Allemandi:2005qs,Nojiri:2006ri}. In the same way, these generalised scalar-torsion $f(T,\phi)$ gravity theories can also be seen as the torsion-based analogue of the so-called generalised $f(R,\phi)$ gravity theories \cite{Faraoni:2004pi,Tsujikawa:2008uc,Alimohammadi:2009yt}, which includes $f(R)$ gravity, scalar field models and scalar-tensor theories, as particular cases. Furthermore, in the context of modified teleparallel gravity, the inclusion of non-linear scalar-torsion coupling terms has been seen as healthy, as it has been shown in Ref. \cite{Gonzalez-Espinoza:2020azh}, they are necessary in order to generate primordial fluctuations during early inflation.  

When we study cosmology in modified gravity theories, we usually obtain complicated systems of equations with ambiguous initial conditions which impedes an analytic treatment, pointing out the necessity of a qualitative analysis using dynamical system theory. This mechanism, known as dynamical analysis, is used to acquire information of the cosmological evolution in the studied system but independently of the initial conditions \cite{amendola2010dark}. Furthermore, although a general cosmological system can exhibit a set of different possible evolutions, its asymptotic behaviour at late-times converges and it is represented by stable critical points obtained from an autonomous system related to the cosmological equations, and the intermediated eras of the cosmological evolution are described by fixed points of the same autonomous system which must be unstable nodes or saddle points \cite{Copeland:2006wr}. In this paper, we use a convenient set of dimensionless variables to study the cosmological dynamics in generalised teleparallel scalar-torsion $f(T, \phi)$ gravity theory. We choose a class of phenomenological model of $f(T, \phi)$ to analyse the critical points and their stability conditions. Particularly, we pay attention to attractors fixed points representing dark energy-dominated solutions, and unstable fixed points which describe scaling matter and scaling radiation eras. To obtain the physical evolution trajectories in the phase space we use the current values of the standard cosmological parameters \cite{Aghanim:2018eyx}, along with the constraints from CMB measurements \cite{Ade:2015rim} and Big Bang Nucleosynthesis (BBN) \cite{Ferreira:1997hj,Bean:2001wt}, applied to the scaling  regimes for early dark energy \cite{Albuquerque:2018ymr,Doran:2006kp}. 

The first time that it was investigated a self-interacting (canonical) scalar field minimally coupled to gravity in the context of $f(T)$ gravity was in Ref. \cite{Yerzhanov:2010vu} where the authors found some analytical solutions that exhibit accelerated expansion. Later in Ref. \cite{Chakrabarti:2017moe}, it was introduced a reconstruction scheme of the function $f(T)$ starting from the scalar potential of a minimally coupled scalar field in an accelerating Universe. On the other hand, in Ref. \cite{Jamil:2012vb} the original framework of a non-minimally coupled scalar field model in teleparallel gravity was extended by replacing $T$ with an arbitrary function $F(T)$. In said reference the authors studied the cosmological dynamics deduced from the associated autonomous system finding that although a rich dynamical behaviour of quintessence and phantom dark energy is observed, no attractor fixed point exists. In this present work we investigate a class of models belonging to a more general scalar-torsion $f(T,\phi)$ gravity theories \cite{Hohmann:2018rwf,Gonzalez-Espinoza:2020azh}, where it is allowed a non-linear gravitational coupling between the scalar field and torsion. Then, the dark energy model constructed from this theory is closely related to the model in Ref. \cite{Jamil:2012vb}, with the difference that here for simplicity we have isolated and kept the pure gravity sector without modifications. The modifications to gravity have been added to the Lagrangian of the scalar field through the generalised non-linear coupling to torsion. For example, this is more similar to what happens in non-minimally coupled scalar field models. Let us emphasize that we have focused on performing a detailed dynamical analysis for this model, studying the attractor and scaling behaviour of its cosmological solutions. We have introduced new ingredients, such as a generalised non-linear coupling to torsion, and then we have obtained a significant progress in the analysis of this kind of models that have not been achieved in the previous studies available in the literature. As a consequence of this, we have found new scaling solutions representing the scaling radiation$/$matter epochs, and new attractors describing the dark energy-dominated era.

The paper is organised as follows. In Section \ref{TG}, we give a brief introduction to TG. In Section \ref{Scalar_Torsion}, we establish the relevant action, and after calculating the background equations, we define the effective dark energy and pressure densities. In Section \ref{Dynamical}, we introduce suitable cosmological variables to write the autonomous system associated with the set of cosmological equations. Thus, after studying in Section \ref{Critical_Points} the critical points of the system, and their stability conditions in Section \ref{Stability}, we perform a numerical treatment for the autonomous system in Section \ref{Numerical}.
Finally, we summarize our findings and our conclusions in Section \ref{Conclu}.


\section{Teleparallel Gravity}\label{TG}

The teleparallel equivalent of General Relativity, or also known as teleparallel gravity (TG) provides an alternative description of gravity in terms of torsion and not curvature \cite{Aldrovandi-Pereira-book,JGPereira2,Arcos:2005ec}. TG is a gauge theory for the translation group \cite{Early-papers5,Early-papers6,Aldrovandi-Pereira-book,Pereira:2019woq}, with the tetrad field $e^{A}_{~\mu}$ playing the role of the dynamical variable of the theory instead the space-time metric $g_{\mu\nu}$, and they are locally related by 
\be
g_{\mu \nu}=\eta_{A B} e^{A}_{~\mu} e^{B}_{~\nu}, 
\ee where $\eta _{AB}^{}=\text{diag}\,(-1,1,1,1)$ is the Minkowski tangent space metric. In a general Lorentz-rotated frame the tetrad field becomes
\be
e^{A}_{~\mu}=\partial_{\mu}{x^{A}}+\omega^{A}_{~B \mu} x^{B}+B^{A}_{~\mu},
\ee
 where the first two terms contain the inertial effects of the frame through the spin connection $\omega^{A}_{~B \mu}$, and the third term $B^{A}_{~\mu}$ is the translational gauge potential representing the gravitational field \cite{Aldrovandi-Pereira-book}. 

The spin connection is defined as
\be
\omega^{A}_{~B \mu}=\Lambda^{A}_{~D}(x) \partial_{\mu}{\Lambda_{B}^{~D}(x)},
\label{spin_TG}
\ee being $\Lambda^{A}_{~D}(x)$ a local (point-dependent) Lorentz transformation. It is a purely inertial spin connection, or flat connection, for which the curvature tensor vanishes identically
\be
R^{A}_{~B \mu\nu}=\partial_{\mu} \omega^{A}_{~B\nu}-\partial_{\nu}{\omega^{A}_{~B \mu}}+\omega^{A}_{~C \mu} \omega^{C}_{~B \nu}-\omega^{A}_{~C \nu} \omega^{C}_{~B \mu}=0.
\ee On the other hand, for a tetrad field that includes the non-trivial translational gauge potential $B^{A}_{~\mu}$ the torsion tensor is non-vanishing and it is given by
\be
T^{A}_{~~\mu \nu}=\partial_{\mu}e^{A}_{~\nu} -\partial_{\nu}e^{A}_{~\mu}+\omega^{A}_{~B\mu}\,e^{B}_{~\nu}
 -\omega^{A}_{~B\nu}\,e^{B}_{~\mu}.
\ee 
The spacetime-indexed linear connection associated with the inertial spin connection $\omega^{A}_{~B\nu}$ is written as
\be
\Gamma^{\rho}_{~~\nu \mu}=e_{A}^{~\rho}\partial_{\mu}e^{A}_{~\nu}+e_{A}^{~\rho}\omega^{A}_{~B \mu} e^{B}_{~\nu},
\ee which is the so-called Weitzenb\"{o}ck connection, and it is related to the Levi-Civita connection $\bar{\Gamma}^{\rho}_{~~\nu \mu}$ through
\be
\Gamma^{\rho}_{~~\nu \mu}=\bar{\Gamma}^{\rho}_{~~\nu \mu}+K^{\rho}_{~~\nu \mu},
\label{RelGamma}
\ee where 
\begin{equation}  \label{Contortion}
 K^{\rho}_{~~\nu\mu}= \frac{1}{2}\left(T^{~\rho}_{\nu~\mu}
 +T^{~\rho}_{\mu~\nu}-T^{\rho}_{~~\nu\mu}\right).
\end{equation} is the contorsion tensor, and
\be
T^{\rho}_{~~\mu \nu}=e_{A}^{~\rho} T^{A}_{~~\mu \nu}=\Gamma^{\rho}_{~~\nu \mu}-\Gamma^{\rho}_{~~\mu \nu}.
\ee is the purely spacetime form of the torsion tensor.

The action of TG is given by \cite{Aldrovandi-Pereira-book}
\be
S=\frac{1}{2 \kappa^2} \int{d^{4}x e ~T},
\ee where $e=\det{(e^{A}_{~\mu})}=\sqrt{-g}$, and $T$ is the torsion scalar that is defined as
\be
T= S_{\rho}^{~~\mu\nu}\,T^{\rho}_{~~\mu\nu},
\label{ScalarT}
 \ee
 where 
\begin{equation} \label{Superpotential}
 S_{\rho}^{~~\mu\nu}=\frac{1}{2}\left(K^{\mu\nu}_{~~~\rho}+\delta^{\mu}_{~\rho} \,T^{\theta\nu}_{~~~\theta}-\delta^{\nu}_{~\rho}\,T^{\theta\mu}_{~~~\theta}\right)\,,
\end{equation} is the so-called super-potential. The gravitational field equations can be obtained by varying with respect to the tetrad field $e^{A}_{~\mu}$, or with respect to $B^{A}_{~\mu}$ . Moreover, using the relation \eqref{RelGamma} one can show that the torsion scalar $T$ and the curvature scalar $R$ of Levi-Civita connection satisfy 
\be
T=-R+2 e^{-1} \partial_{\mu}(e T^{\nu \mu}_{~~~\nu}),
\label{Equiv} 
\ee and therefore, TG and GR are equivalent at the level of field equations. 

However, in the same way as one can modify gravity starting from GR, one can also modify gravity starting from TG, either by introducing a non-minimally coupled matter field, as for example a scalar field \cite{Geng:2011aj,Otalora:2013tba,Otalora:2013dsa,Otalora:2014aoa,Skugoreva:2014ena}, or adding into the action non-linear terms in the torsion scalar $T$, as for example in $f(T)$ gravity \cite{Bengochea:2008gz,Linder:2010py,Li:2011wu,Gonzalez-Espinoza:2018gyl}. In all the cases, because the relation \eqref{Equiv} only guarantees the equivalence with GR for a gravitational action linear in torsion or decoupled from other fields, we obtain new classes of modified gravity theories not equivalent to their corresponding counterpart based on curvature. Furthermore, it has been seen that these gravitational modifications based on torsion  have a rich phenomenology which has resulted in a fair number of articles in cosmology of early and late-time Universe \cite{Cai:2015emx}.

Below we investigate the dynamics of a class of scalar-torsion $f(T,\phi)$ gravity theories \cite{Hohmann:2018rwf,Gonzalez-Espinoza:2020azh} that include features from both non-minimally coupled scalar field models, and modify gravity models with non-linear mater-gravity coupling \cite{Harko:2014aja,Harko:2014sja,Carloni:2015lsa,Gonzalez-Espinoza:2018gyl}, inspired in the similar constructions based on curvature such as in non-minimally coupled $f(R)$ gravity models \cite{Nojiri:2004bi,Allemandi:2005qs,Nojiri:2006ri,Bertolami:2007gv,Harko:2008qz,Harko:2010mv,Bertolami:2009ic,Bertolami:2013kca,Wang:2013fja}, and the so-called generalised $f(R,\phi)$ gravity theories \cite{Faraoni:2004pi,Tsujikawa:2008uc,Alimohammadi:2009yt}.


\section{Scalar-Torsion $f(T,\phi)$ Gravity}\label{Scalar_Torsion}

The relevant action is \cite{Hohmann:2018rwf,Gonzalez-Espinoza:2020azh}
\begin{equation}
 S=\int d^{4}x\,e\,\left[ f(T,\phi)+ P(\phi)X \right]+S_{m}+S_{r},
\label{action_Scalar_Torsion}
\end{equation} where $f(T,\phi)$ is an arbitrary function of the torsion scalar $T$ and a scalar field $\phi$ and $X=-\partial_{\mu}{\phi}\partial^{\mu}{\phi}/2$. Also, $S_{m}$ is the action of non-relativistic matter, including baryons and dark matter, and $S_{r}$ is the action describing the radiation component. 

In choosing the cosmological background, we assume the diagonal tetrad field
\begin{equation}
\label{veirbFRW}
e^A_{~\mu}={\rm
diag}(1,a,a,a),
\end{equation} which is a proper tetrad naturally associated with the vanishing spin connections $\omega^{A}_{~ B\mu}=0$ \cite{Krssak:2015oua}, and which leads to the flat Friedmann-Lema\^{i}tre-Robertson-Walker (FLRW) metric
\begin{equation}
ds^2=-dt^2+a^2\,\delta_{ij} dx^i dx^j \,,
\label{FRWMetric}
\end{equation} where $a$ is the scale factor, function of the cosmic time $t$. Hence, the background equations are given by
\bea
\label{00}
  f(T,\phi) - P(\phi) X - 2 T f_{,T}=\rho_m &+& \rho_r, \ \ \ \ \ \\
\label{ii}
 f(T,\phi) + P(\phi) X - 2 T f_{,T} - 4 \dot{H} f_{,T} - 4 H \dot{f}_{,T} =&-&p_r, \\
\label{phi}
 - P_{,\phi} X - 3 P(\phi) H \dot{\phi} - P(\phi) \ddot{\phi} + f_{,\phi}&=&0,
\eea
where $H\equiv \dot{a}/a$ is the Hubble rate, a dot represents derivative with respect to $t$, and a comma denotes derivative with respect to $\phi$ or $T$. Also, the functions $\rho_{i}$, $p_{i}$, with $i=m,r$ are the energy and pressure densities of non-relativistic matter (cold dark matter and baryons), and radiation, respectively, being that we have already used in the above equations the corresponding barotropic equations of state $w_{m}=p_{m}/\rho_{m}=0$ and $w_{r}= p_{r}/\rho_{r}=1/3$.  

In order to proceed forward we are going to consider the class of models with \cite{Gonzalez-Espinoza:2020azh}
\be
f(T,\phi)=-\frac{1}{2 \kappa^2} T-F(\phi) G(T) -V(\phi), 
\ee where $V(\phi)$ is the scalar potential, $F(\phi)$ the non-minimal coupling function of $\phi$, and $G(T)$ an arbitrary function of $T$. Then, we obtain
\bea
\label{H00}
\frac{3}{\kappa^2} H^2 & = & G(T) F(\phi) - 2 T G_{,T} F(\phi) + V + P(\phi) X \nonumber\\
&& + \rho_{m}+\rho_{r},\\
-\frac{2}{\kappa^2} \dot{H} & = & 2 P(\phi) X + 4 \dot{H} G_{,T} F(\phi) + 4 H G_{,TT} \dot{T} F(\phi) \nonumber\\
&& + 4 H G_{,T} \dot{F} + \rho_{m} + \frac{4}{3}\rho_{r},
\label{Hii}
\eea
while the motion equation of $\phi$ is written as 
\begin{equation}
P(\phi) \ddot{\phi} + 3 P(\phi) H \dot{\phi} + P_{,\phi} X + G(T) F_{,\phi}+ V_{,\phi}=0.
\end{equation}

Following Ref. \cite{Copeland:2006wr}, the Friedmann equations \eqref{H00} and \eqref{Hii} can also be rewritten as 
\bea
\label{SH00}
&& \frac{3}{\kappa^2} H^2=\rho_{de}+\rho_{m}+\rho_{r},\\
&& -\frac{2}{\kappa^2} \dot{H}=\rho_{de}+p_{de}+\rho_{m}+\frac{4}{3}\rho_{r},
\label{SHii}
\eea where we have defined the energy and pressure densities of dark energy in the way
\bea
\label{rho_de}
 \rho_{de}&=& P(\phi) X + V + (G-2 T G_{,T}) F(\phi), \\
 p_{de}&=&P(\phi) X - V -(G-2 T G_{,T}) F(\phi)\nonumber\\
&& + 4 ( 2 T G_{,TT} + G_{,T} ) F(\phi) \dot{H} + 4 H G_{,T} F_{,\phi}\dot{\phi}.
\label{p_de}
\eea
Furthermore, we can define the effective dark energy equation-of-state parameter as
\begin{equation}
w_{de}=\frac{p_{de}}{\rho _{de}}.
\label{wDE1}
\end{equation}
One can easily see that $\rho_{de}$ and $p_{de}$ obey the standard evolution equation 
\begin{eqnarray}
\dot{\rho}_{de}+3H(\rho_{de}+p_{de})=0.
\end{eqnarray} which is consistent with energy conservation law and the fluid evolution equations
\bea
\label{rho_m}
&& \dot{\rho}_{m}+3 H\rho_{m}=0,\\
&& \dot{\rho}_{r}+4 H\rho_{r}=0.
\label{rho_r}
\eea
Lastly, concerning with the cosmological investigations, it proves convenient to introduce the 
total equation-of-state parameter as 
\be
w_{tot}=\frac{p_{de}+p_r}{\rho _{de}+\rho _m+\rho _r},
\label{wtot}
\ee
which is immediately related to the deceleration parameter $q$ through 
\be
q=\frac{1}
{2}\left(1+3w_{tot}\right),
\label{deccelparam}
\ee
and hence acceleration occurs when $q<0$, as well as the standard density 
parameters  
\be
\Omega_{m}\equiv\frac{\kappa^2 \rho_{m}}{3 H^2},\:\:\:\: \Omega_{de}\equiv\frac{\kappa^2 \rho_{de}}{3 H^2},\:\:\:\: \Omega_{r}\equiv \frac{\kappa^2 \rho_{r}}{3 H^2},
\ee such that 
\be
\Omega_{de}+\Omega_{m}+\Omega_{r}=1.
\ee

In order to find cosmological solutions and to study the complete dynamics in the phase space for this class of dark energy
models we are going to assume the ansatz 
\be
G(T)=\left(\frac{T}{6}\right)^{1+s},
\ee and $P=1$. This expression is inspired by modify gravity models with non-linear matter-gravity coupling \cite{Nojiri:2004bi,Allemandi:2005qs,Nojiri:2006ri}, but neglecting the kinetic term of the scalar field. Including the kinetic term could deviate the squared tensor propagation speed from $1$ \cite{Gonzalez-Espinoza:2019ajd}, which is something undesirable \cite{Baker:2017hug,Sakstein:2017xjx}. This non-linear scalar-torsion coupling is also motivated from the physics of the very early universe, where it is associated with the generation of primordial fluctuations during inflation, in the context of $f(T,\phi)$ gravity \cite{Gonzalez-Espinoza:2020azh}.

In this case, the energy and pressure densities of dark energy can be written as
\bea
\label{rho_de_model}
 \rho_{de}&=& \dfrac{\dot{\phi}^2}{2} + V - (1+2 s) H^{2 (1+s)} F(\phi), \\
 p_{de}&=& \dfrac{\dot{\phi}^2}{2} - V + (1+2 s)  H^{2 (1+s)} F(\phi)\nonumber\\
&& + \dfrac{2}{3} (1+s) (1+2 s) H^{2 s} F(\phi) \dot{H} \nonumber\\
&& + \dfrac{2}{3} (1+s)   H^{1+2 s} F_{,\phi}\dot{\phi},
\label{p_de_model}
\eea and the motion equation of $\phi$ becomes
\begin{equation}
\ddot{\phi} + 3 H \dot{\phi} + H^{2 (1+s)} F_{,\phi}+ V_{,\phi}=0.
\label{Eq_phi_model}
\end{equation}

Thus, Eqs. \eqref{SH00}, \eqref{SHii}, along with equations \eqref{rho_de_model}, \eqref{p_de_model}, and Eqs. \eqref{rho_m}, \eqref{rho_r}, and \eqref{Eq_phi_model}, compose the set of cosmological equations for the model.

\section{Dynamical system}\label{Dynamical}

To obtain the corresponding autonomous system associated with the above set of cosmological equations, we introduce the following useful dimensionless variables \cite{Copeland:2006wr,Bahamonde:2017ize}
\bea
&& x= \dfrac{\kappa \dot{\phi}}{\sqrt{6}H}, \:\:\: y= \dfrac{\kappa \sqrt{V}}{\sqrt{3} H}, \nonumber\\
&& u =-\frac{1}{3}\kappa^2 (2 s+1) F(\phi) H^{2 s}, \:\:\: {\Varrho} =\frac{\kappa \sqrt{\rho_{r}}}{\sqrt{3} H},
\label{Var}
\eea
and
\bea
&& \lambda= -\dfrac{V'(\phi)}{V(\phi)},\:\:\:\: \sigma= -\dfrac{F'(\phi)}{F(\phi)},\\
&& \Gamma= \dfrac{V(\phi)V''(\phi)}{V'(\phi)^2},\:\:\:\: \Theta= \dfrac{F(\phi)F''(\phi)}{F'(\phi)^2},
\eea
that satisfy the constraint equation
\begin{equation}
x^2+y^2+u+\Omega_m+{\Varrho}^2=1.
\end{equation} 

Therefore, we obtain the dynamical system

\begin{eqnarray}
\label{ODE1}
\dfrac{\text{d}x}{\text{d}N} &=& \dfrac{- f_1(x,y,u,{\Varrho})}{2(1+2 s)\left[(s+1) u-1\right]},\\
\dfrac{\text{d}y}{\text{d}N} &=& \dfrac{-y f_2(x,y,u,{\Varrho})}{2(1+2 s)\left[(s+1) u-1\right]},\\
\label{ODE2}
\dfrac{\text{d}u}{\text{d}N} &=& \dfrac{u f_3(x,y,u,{\Varrho})}{(1+2 s)\left[(s+1) u-1\right]},\\
\label{ODE3}
\dfrac{\text{d}{\Varrho}}{\text{d}N}&=& \dfrac{-{\Varrho} f_4(x,y,u,{\Varrho})}{2 (1+2 s)\left[(s+1) u-1\right]},\\
\label{ODE4}
\frac{\text{d}\lambda}{\text{d}N}&=& -\sqrt{6} (\Gamma -1) \lambda ^2 x,\\
\label{ODE5}
\frac{\text{d}\sigma}{\text{d}N}&=& -\sqrt{6} (\Theta -1) \sigma ^2 x,
\label{ODE6}
\end{eqnarray}
where we have defined
\bea
 f_1(x,y,u,{\Varrho})&=&(6 s+3) x^3+2 \sqrt{6} (s+1) \sigma  x^2 u+\nonumber\\
&& (2 s+1) x \left[3(2 s+1) u-3 y^2+{\Varrho} ^2-3\right]+\nonumber\\
&& \sqrt{6}\left[(s+1) u-1\right] \left[\sigma  u-\lambda y^2 (1 +2 s)\right],\\
f_2(x,y,u,{\Varrho})&=&\sqrt{6} x \left[(s+1) u \left[\lambda\left(1+2 s\right)+2 \sigma \right]-\lambda  (2 s+1)\right]-\nonumber\\
&& (2 s+1) \left[3 \left(y^2+u-1\right)-{\Varrho} ^2\right]+ \nonumber\\
&&(6 s+3) x^2,\\
f_3(x,y,u,{\Varrho})&=&3 s (2 s+1) x^2-\sqrt{6} \sigma  x \left[(s+1) u-2 s-1\right]+\nonumber\\
&& s (2 s+1) \left[{\Varrho} ^2-3 \left(y^2+u-1\right)\right],\\
f_4(x,y,u,{\Varrho})&=&3 (2 s+1) x^2+2 \sqrt{6} (s+1) \sigma  x u+\nonumber\\
&& (2 s+1) \left(4 s u-3 y^2+u+{\Varrho} ^2-1\right).
\eea 

Using the above set of phase space variables we can also write 
\be
\Omega_{de}=x^2+y^2+u,\:\:\: \Omega_{m}=1-x^2-y^2-u-{\Varrho}^2,\:\:\: \Omega_{r}= {\Varrho}^2.
\ee 

Similarly, the equation of state of dark energy $w_{de}=p_{de}/\rho_{de}$ can be rewritten as
\bea
w_{de} &=& \frac{2 \sqrt{\frac{2}{3}} \sigma  x u}{(u-2) \left[u\left(s+1\right)-1\right] \left(x^2+y^2+u\right)}-\nonumber\\
&& \frac{\dfrac{2 \sqrt{6} \sigma  x u}{(2 s+1) (u-2)}+{\Varrho} ^2+3}{3 \left(x^2+y^2+u\right)}-\nonumber\\
&&\frac{3 x^2-3 \left(y^2+u-1\right)+{\Varrho} ^2}{3 \left[u(s+1)-1\right] \left(x^2+y^2+u\right)},
\eea
whereas the total equation of state becomes
\bea
w_{T}&=&-1+ \frac{1}{s+1}+ \frac{(s+1)\left( y^2 - x^2\right)-s}{(s+1) \left[u(s+1)-1\right]}+\\
&& \frac{4 \sqrt{\frac{2}{3}} \sigma  x}{(2 s+1) (u-2)}-\frac{(2 s+1){\Varrho} ^2+2 \sqrt{6} \sigma  x}{3 (2 s+1) \left[(s+1) u-1\right]}-\nonumber\\
&& -\frac{2 \sqrt{\frac{2}{3}} \sigma  x u}{(2 s+1) (u-2)}.
\eea
The dynamical system \eqref{ODE1}-\eqref{ODE6} is not an autonomous system unless the parameters $\Gamma$ and $\Theta$ are known \cite{Copeland:2006wr,Bahamonde:2017ize}. From now we concentrate in the exponential potential $V(\phi)=V_{0} e^{-\lambda\kappa \phi}$, with $\lambda$ a dimensionless constant, that is, $\Gamma=1$. Let us remember that this scalar potential can give rise to an accelerated expansion, and at the same time, it allows to obtain scaling solutions \cite{amendola2010dark,Copeland:2006wr}. On the other hand, for the non-minimal coupling function of $\phi$ we take $F(\phi)=F_{0}e^{-\sigma \kappa \phi}$, such that $\Theta=1$. This is the most natural and simple choice compatible with the exponential scalar potential \cite{Amendola:1999qq}.  

\section{Critical points}\label{Critical_Points}

We obtain the critical points or fixed points $(x_{c},y_{c},u_{c},{\Varrho}_{c})$ of the corresponding autonomous system by imposing the conditions $dx/dN=dy/dN=du/dN=d{\Varrho}/dN=0$. From the definition \eqref{Var}, the values $x_{c}$,$y_{c}$, $u_{c}$ and ${\Varrho}_{c}$ must be reals with $y_{c}\geq 0$ and ${\Varrho}_{c}\geq 0$. The critical points are presented in the Table \ref{table1}, while the expressions for the cosmological parameters for each critical point are shown in Table \ref{table2}.

The point $a_{R}$ is a radiation-dominated solution $\Omega_{r}=1$ with a total equation of state $w_{tot}=1/3$. The equation of state of dark energy takes the value $w_{de}=(4 s+1)/3$, which depends on the parameter $s$. It exists for all the values of parameters $s$, $\sigma$ and $\lambda$. Point $b_{R}$ is a scaling solution with $\Omega_{de}^{(r)}=4/\lambda^2$, and $\Omega_{m}=0$. For this point we found $w_{de}=w_{tot}=1/3$. The physical condition $0<\Omega_{de}^{(r)}<1$ imposes the constraint $\abs{\lambda}>2$. Point $c_{R}$ is also a scaling solution which is a new solution that exist only for $s\neq 0$. The fractional energy density parameter of dark energy is $\Omega_{de}^{(r)}=4 s (4 s+1)/(3 \sigma ^2)$, and $\Omega_{m}=0$. The constraints for the parameters due to the physical condition $0<\Omega_{de}^{(r)}<1$ have been put in Table \ref{table3} . This point also satisfies $w_{de}=w_{tot}=1/3$. So, points $b_{R}$ and $c_{R}$ describe a non-standard radiation-dominated era in which there is a small contribution coming from dark energy. Thus, if $b_{R}$ and  $c_{R}$ are both responsible for the scaling radiation era, we also need  to consider the earliest constraint coming from physics of big bang nucleosynthesis (BBN) which requires $\Omega_{de}^{(r)}<0.045$  \cite{Ferreira:1997hj,Bean:2001wt}. So, in the case of point $b_{R}$ we find $\abs{\lambda}>9.94$, while for $c_{R}$ we obtain $-0.25<s\leq 0$ for $\sigma \neq 0$, or, $s\leq -0.25$ (or $s>0$) and $\abs{\sigma}> 5.44 \sqrt{s \left(4 s+1\right)}$.

Point $d_{M}$ represents a standard cold dark matter-dominated era with $\Omega_{m}=1$, $w_{de}=s$ and $w_{tot}=0$. This point exists for all the values of parameters. Points $e$ and $f$ satisfy $\Omega_{de}=1$, but they cannot explain the current accelerated expansion because them behave as stiff matter with $w_{de}=w_{tot}=1$. On the other hand, point $g_{M}$ describes a non-standard cold dark matter-dominated era with a small contribution of the fractional dark energy density parameter given by $\Omega_{de}^{(m)}=3/\lambda ^2$, $\Omega_{r}=0$, and $w_{de}=w_{tot}=0$. The point $i_{M}$ is a new fixed point which is present only for $s\neq 0$ and it is a scaling solution representing a non-standard cold dark matter-dominated era with $\Omega_{de}^{(m)}=3 s (3 s+1)/(2 \sigma ^2)$, $\Omega_{r}=0$, and $w_{de}=w_{tot}=0$. The constraints for the parameters obtained from the physical condition $0<\Omega_{de}^{(m)}<1$ are shown in Table \ref{table3}.  If points $g_{M}$ and $i_{M}$ represent both the scaling matter era they are constrained to satisfy $\Omega_{de}^{(m)}<0.02$ ($95\%$ C.L.), at redshift $z\approx 50$, according to CMB measurements \cite{Ade:2015rim}. Thus, for $g_{M}$ we find $\abs{\lambda}>12.25$, while for $i_{M}$ we get $-0.\bar{3}<s\leq 0$ for $\sigma \neq 0$, or $s\leq -0.\bar{3}$ (or $s>0$) and $\abs{\sigma}> \sqrt{s (225 s+75)}$.

Point $h$ is identified as dark energy-dominated era with $\Omega_{de}=1$, and $w_{de}=w_{tot}=(\lambda^2-3)/3$. It exists for $\abs{\lambda}<\sqrt{6}$ and it can explain the current cosmic acceleration for $\abs{\lambda}<\sqrt{2}$ \cite{Copeland:2006wr}. 
Points $j$ and $k$ provide dark energy-dominated eras which can explain the cosmic accelerated expansion. These are also new solutions of dark energy which are present only for $s\neq 0$. The expressions for $w_{de}$ and $w_{tot}$ are shown in Table \ref{table2}, whereas the existence and accelerated expansion conditions are detailed in Table \ref{table3}. Finally, point $l$ is a de Sitter solution with $\Omega_{de}=1$, and $w_{de}=w_{tot}=-1$, which provides accelerated expansion for the all values of the parameters. Although this points exist for $s=0$, the new expression for the phase space coordinate $y_{c}$ associated with it and reported in Table \ref{table1} is a generalisation of the case $s=0$, which now also includes values for $s\neq 0$. The conditions of existence for this point have also been detailed in Table \ref{table3}.

\begin{table*}
 \centering
 \caption{Critical points for the autonomous system  \eqref{ODE1}-\eqref{ODE6} for   $V(\phi)=V_{0} e^{-\lambda\kappa \phi}$ and $F(\phi)=F_{0}e^{-\sigma \kappa \phi}$. We define $A=9 s^2 (2 s+1)^2-6 s (s+1) \sigma ^2$.}
\begin{center}
\begin{tabular}{c c c c c c c c c}\hline\hline
Name &  $x_{c}$ & $y_{c}$ & $u_{c}$ & ${\Varrho}_{c}$  \\\hline
$\ \ \ \ \ \ \ \ a_{R} \ \ \ \ \ \ \ \ $ & $0$  & $0$  & $0$ & $1$   \\
$b_{R}$ & $\frac{2 \sqrt{\frac{2}{3}}}{\lambda }$  & $\frac{2}{\sqrt{3} \lambda }$  & $0$ & $\sqrt{1-\frac{4}{\lambda ^2}}$   \\
$c_{R}$ & $-\frac{2 \sqrt{\frac{2}{3}} s}{\sigma }$  & $0$  & $\frac{4 s (2 s+1)}{3 \sigma ^2}$ & \ \ \ \ \ $\sqrt{1-\frac{4 s (4 s+1)}{3 \sigma ^2}}$ \ \ \ \ \   \\
$d_{M}$ & $0$  & $0$  & $0$ & $0$  \\
$e$ & $-1$  & $0$  & $0$ & $0$   \\
$f$ & $1$  & $0$  & $0$ & $0$   \\
$g_{M}$ & $\frac{\sqrt{\frac{3}{2}}}{\lambda }$  & $\sqrt{\frac{3}{2}} \sqrt{\frac{1}{\lambda ^2}}$  & $0$ & $0$   \\
$h$ & $\frac{\lambda }{\sqrt{6}}$  & $\sqrt{1-\frac{\lambda ^2}{6}}$  & $0$ & $0$   \\
$i_{M}$ & $-\frac{\sqrt{\frac{3}{2}} s}{\sigma }$  & $0$  & $\frac{3 s (2 s+1)}{2 \sigma ^2}$ & $0$   \\
$j$ & $-\frac{3 s (2 s+1)+\sqrt{A}}{\sqrt{6} (s+1) \sigma }$  & $0$  & $\frac{(2 s+1) \left(3 s \left(x_{c}^2+1\right)+\sqrt{6} \sigma  x_{c}\right)}{\sqrt{6} (s+1) \sigma  x_{c}+3 s (2 s+1)}$ & $0$   \\
$k$ & $\frac{-3 s (2 s+1)+\sqrt{A}}{\sqrt{6} (s+1) \sigma }$  & $0$  & $\frac{(2 s+1) \left(3 s \left(x_{c}^2+1\right)+\sqrt{6} \sigma  x_{c}\right)}{\sqrt{6} (s+1) \sigma  x_{c}+3 s (2 s+1)}$ & $0$   \\
$l$ & $0$  & \ \ \ \ \ $\sqrt{\frac{\sigma }{(2 s+1) \left(\lambda +\frac{\sigma }{2 s+1}\right)}}$ \ \ \ \ \  & $\frac{\lambda  (2 s+1)}{\lambda +2 \lambda  s+\sigma }$ & $0$   \\
\\ \hline\hline
\end{tabular}
\end{center}
\label{table1}
\end{table*}

\begin{table}
 \centering
 \caption{Cosmological parameters for the critical points in Table \ref{table1}. We define $A=9 s^2 (2 s+1)^2-6 s (s+1) \sigma ^2$. The fractional energy density of the radiation fluid is calculated through $\Omega_{r}={\Varrho}^2=1-\Omega_{m}-\Omega_{de}$.}
\begin{center}
\begin{tabular}{c c c c c}\hline\hline
Name &   $\Omega_{de}$ & $\Omega_{m}$ & $w_{de}$ & $w_{tot}$ \\\hline
$a_{R}$ & $0$ & $0$ & $\frac{1}{3} (4 s+1)$ & $\frac{1}{3}$ \\
$b_{R}$ &  $\frac{4}{\lambda ^2}$ & $0$ & $\frac{1}{3}$ & $\frac{1}{3}$ \\
$c_{R}$ &  $\frac{4 s (4 s+1)}{3 \sigma ^2}$ & $0$ & $\frac{1}{3}$ & $\frac{1}{3}$ \\
$d_{M}$ &  $0$ & $1$ & $s$ & $0$ \\
$e$ &  $1$ & $0$ & $1$ & $1$ \\
$f$ &  $1$ & $0$ & $1$ & $1$ \\
$g_{M}$ &  $\frac{3}{\lambda ^2}$ & $1-\frac{3}{\lambda ^2}$ & $0$ & $0$ \\
$h$ &  $1$  & $0$ & $\frac{1}{3} \left(\lambda ^2-3\right)$ & $\frac{1}{3} \left(\lambda ^2-3\right)$ \\
$i_{M}$ &  $\frac{3 s (3 s+1)}{2 \sigma ^2}$ & $1-\frac{3 s (3 s+1)}{2 \sigma ^2}$ & $0$ & $0$ \\
$j$ &  $1$   & $0$ & $\frac{\sqrt{A}+3 s^2}{3 s (s+1)}$ & $\frac{\sqrt{A}+3 s^2}{3 s (s+1)}$ \\
$k$ &  $1$  & $0$ & $\frac{3 s^2-\sqrt{A}}{3 s (s+1)}$ & $\frac{3 s^2-\sqrt{A}}{3 s (s+1)}$ \\
$l$ &  $1$  & $0$ & $-1$ & $-1$ \\
\\ \hline\hline
\end{tabular}
\end{center}
\label{table2}
\end{table}

\section{Stability of critical points}\label{Stability}

In order to study the stability of the critical points we consider time dependent liner perturbations $\delta{x}$, $\delta{y}$, $\delta{u}$ and $\delta{\Varrho}$ around each critical point in the form $x=x_{c}+\delta{x}$, $y=y_{c}+\delta{y}$, $u=u_{c}+\delta{u}$, and ${\Varrho}={\Varrho}_{c}+\delta{\Varrho}$. By substituting these expressions into the autonomous system \eqref{ODE1}-\eqref{ODE4} and linearising them we obtain the linear perturbation matrix $\mathcal{M}$ \cite{Copeland:2006wr}. The eigenvalues of $\mathcal{M}$, namely, $\mu_{1}$, $\mu_{2}$, $\mu_{3}$ and $\mu_{4}$ evaluated at each fixed point determines the stability conditions for each one of them. The classification of the stability properties is established usually in the following way: (i) Stable node: all the eigenvalues are negative;
(ii) Unstable node: all the eigenvalues are positive; (iii) Saddle point: one, two, or three of the four eigenvalues are positive and the others are negative; (iv) Stable spiral: The determinant of $\mathcal{M}$ is negative, and the real part of all the eigenvalues are negative. Points which are stable node or stable spiral are called attractor points, and these fixed points are reached through the cosmological evolution of the Universe, independently of the initial conditions. The conditions of existence, stability and acceleration of critical points for system \eqref{ODE1}-\eqref{ODE4} are shown in Table \ref{table3}.

\begin{itemize}

\item[•]
Point $a_{R}$ has the eigenvalues
\be
\mu_{1}=-1,\:\:\:\mu_{2}=1,\:\:\: \mu_{3}=2,\:\:\:\mu_{4}=-4s,
\ee then this point is always a saddle point. \\

\item[•]
For point $b_{R}$ we obtain
\bea
&& \mu_{1}=1,\:\:\: \mu_{2,3}=\frac{1}{2} \left(-1\mp\sqrt{\frac{64}{\lambda ^2}-15}\right),\nonumber\\
&& \mu_{4}=-\frac{4 (\lambda  s+\sigma )}{\lambda },
\eea which implies that it is a saddle point for all the values of parameters. \\

\item[•]
Similarly, point $c_{R}$ has the eigenvalues
\bea
\mu_{1}&=&1,\:\:\: \mu_{2}=\frac{2 \lambda  s}{\sigma }+2,\:\:\: \nonumber\\
\mu_{3,4}&=&-\frac{1}{2}\mp \sqrt{\frac{4 s \left[4 s (4 s+1)-3 \sigma ^2\right]}{4 s (s+1) (2 s+1)-3 \sigma ^2}+\frac{1}{4}},
\eea and it is also always a saddle point. \\

\item[•]
Point $d_{M}$ leads us to
\bea
\mu_{1}=-\frac{1}{2},\:\:\:\mu_{2}=-\frac{3}{2},\:\:\: \mu_{3}=\frac{3}{2},\:\:\:\mu_{4}=-3s.
\eea which tells us that it is always a saddle point. \\

\item[•]
For points $e$ and $f$ we find 
\be
\mu_{1}=3,\:\:\:\mu_{2}=1,\:\:\: \mu_{3}=3\pm\sqrt{\frac{3}{2}} \lambda,\:\:\:\mu_{4}= -6 s\pm\sqrt{6} \sigma,
\ee where $(+)$ corresponds to $e$ and $(-)$ to $f$. Point $e$ is an unstable node for $s<\frac{\sigma }{\sqrt{6}}$ and $\lambda >-\sqrt{6}$ with $\sigma \in \mathbb{R}$. In any other case it is a saddle point. Also, point $f$ is an unstable node for $\lambda <\sqrt{6}$ and $s<-\frac{\sigma }{\sqrt{6}}$, with $\sigma \in \mathbb{R}$. If one of these conditions is not satisfied then it is a saddle point. \\

\item[•]
Point $g_{M}$ has associated the eigenvalues
\bea
&& \mu_{1}=-\frac{1}{2},\:\:\:\mu_{2,3}=\frac{3}{4} \left(-1\mp\sqrt{\frac{24}{\lambda ^2}-7}\right),\nonumber\\
&& \mu_{4}=-\frac{3 (\lambda  s+\sigma )}{\lambda },
\eea which means that it is a saddle point for $-2 \sqrt{\frac{6}{7}}\leq \lambda <-\sqrt{3}$ and $\sigma >-\lambda s$,  or $\sqrt{3}<\lambda \leq 2 \sqrt{\frac{6}{7}}$ and $\sigma <-\lambda s$, with $s \in \mathbb{R}$.  On the other hand, it is a stable node for $s\in \mathbb{R}$, and, $-2 \sqrt{\frac{6}{7}}<\lambda <-\sqrt{3}\land \sigma <-\lambda s$, or $\sqrt{3}<\lambda <2 \sqrt{\frac{6}{7}}\land \sigma >-\lambda s$. However, this point cannot provide the current accelerated expansion of the Universe. \\

\item[•]
For point $h$ we get 
\bea
\mu_{1}&=&\frac{1}{2} \left(\lambda ^2-6\right),\:\:\:\mu_{2}=\frac{1}{2} \left(\lambda ^2-4\right),\:\:\: \nonumber\\
\mu_{3}&=&\lambda ^2-3,\:\:\: \mu_{4}=-\lambda  (\lambda  s+\sigma ).
\eea This critical point has a range in the parameters space with accelerated expansion, and we are interested in its stability conditions within this range. One finds that when point $h$ has accelerated expansion it is a stable node for $-\sqrt{2}<\lambda <0\land \sigma <-\lambda s$, or $0<\lambda <\sqrt{2}\land \sigma >-\lambda s$, with $s \in \mathbb{R}$, and in other cases it is a saddle point. \\

\item[•]
In the case of point $i_{M}$ one finds
\bea
\mu_{1}&=&-\frac{1}{2},\:\:\:\mu_{2}=\frac{3 (\lambda  s+\sigma )}{2 \sigma },\:\:\:  \nonumber\\
\mu_{3,4}&=&-\frac{3}{4} \mp 3\sqrt{\frac{3 s^4}{2 \sigma ^2-3 s (s+1) (2 s+1)}+\frac{s}{2}+\frac{1}{16}}. \nonumber\\
&&
\eea Since it is a matter solution we are interested in unstable regions of the parameters space. We obtain a saddle point, which is always unstable, with the corresponding region of parameters in table \ref{table3}. Oppositely, this scaling solution can also be a stable node, as for example for $s>0 \ \land \ -\sqrt{3} \sqrt{\frac{s (s (26 s+11)+1)}{16 s+2}}$ $<\sigma <-\sqrt{\frac{3}{2}} \sqrt{s (3 s+1)}\land \lambda >-\frac{\sigma }{s}$. Nonetheless, it is not viable to explain a late-time acceleration. \\

\item[•]
Also, for points $j$ and $k$ we obtain
\bea
&& \mu_{1}=\frac{\left[3 s (2 s+1)\pm\sqrt{A}\right] (\lambda  s+\sigma )}{2 s (s+1) \sigma },\:\:\:\mu_{2}=\frac{-3 s\pm\sqrt{A}}{2 s (s+1)},\:\:\: \nonumber\\
&& \mu_{3}=\frac{s (2 s-1)\pm\sqrt{A}}{2 s (s+1)}, \mu_{4}=\frac{3 s^2\pm\sqrt{A}}{s (s+1)},
\eea where $A=9 s^2 (2 s+1)^2-6 s (s+1) \sigma ^2$, and sign $(+)$ is for $j$, and $(-)$ for $k$. Both points are dark energy solutions and they can explain the current accelerated expansion of the Universe. Therefore we are interested in finding the stability conditions for these points when them provide accelerated expansion. From the above eigenvalues we find that exists a region of the space of parameters in which these points are stable nodes and thus attractors. These constraints for the parameters are shown in Table \ref{table3}. \\

\item[•]
Finally, for point $l$ we find the eigenvalues
\bea
&& \mu_{1}=-2,\:\:\:\mu_{2}=-3,\nonumber\\
&& \mu_{3,4}=-\frac{3}{2}\mp\sqrt{\frac{9}{4}-\frac{3 \lambda  \sigma  (\lambda  s+\sigma )}{\lambda  s (2 s+1)-\sigma }}.
\eea This is a de Sitter solution which therefore provides accelerated expansion for all the values of the parameters. We find that it is a stable node for $\lambda >0$  $\land$  $s>0$  $\land$
\begin{eqnarray*}
&&0<\sigma \leq \frac{\left[-4 \lambda ^2 s+\lambda  \sqrt{\frac{9}{\lambda ^2}+8 s \left(2 \left(\lambda ^2+6\right) s+9\right)}-3\right]}{8 \lambda} .
\end{eqnarray*}
Also, for $s<0$ this point can be a stable node under some constraints of the parameters $s$, $\lambda$ and $\sigma$, but the expressions for these constraints are extremely large and then we do not put them explicitly here. Finally, this point is never a stable spiral. 

\end{itemize}

\begin{figure}[htbp]
	\centering
		\includegraphics[width=0.4\textwidth]{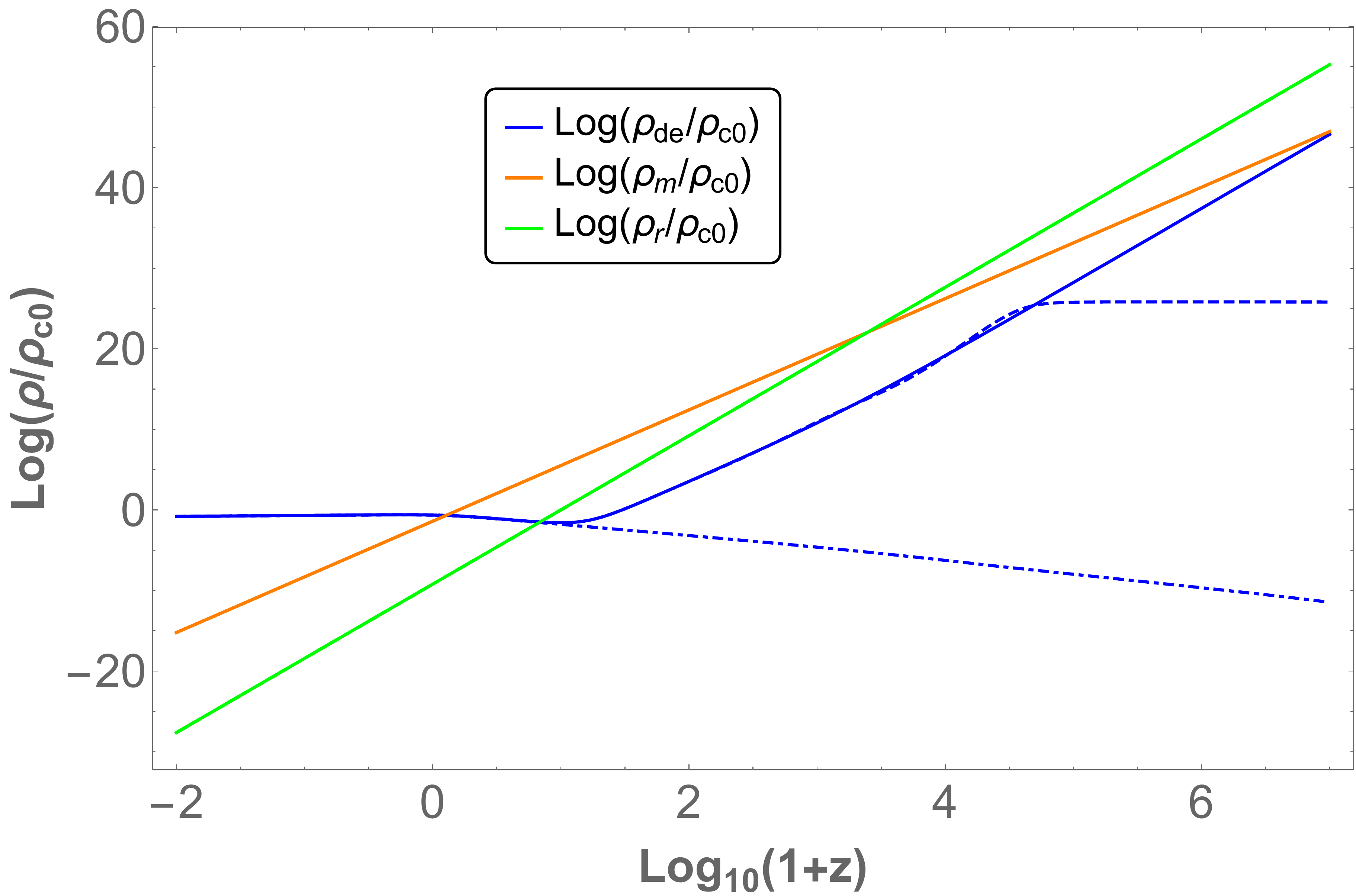}
	\caption{We depict the evolution of the energy density of dark energy $\rho_{de}$ (blue), dark matter (including baryons) $\rho_{m}$ (orange) and radiation $\rho_{r}$ (green) as functions of the redshift $z$, for $s=-1.20$, $\sigma=0.3$ and $\lambda=150$. The solid blue line corresponds to the initial conditions $x_{i}=0.0108$, $y_{i}=0.00765$, $u_{i}=9.3\times 10^{-30}$, ${\Varrho}_{i}=0.999788$, dashed blue line to $x_{i}=1\times 10^{-9}$, $y_{i}=4\times 10^{-7}$, $u_{i}=9.1\times 10^{-30}$, ${\Varrho}_{i}=0.999877$, and dot-dashed blue line to $x_{i}=1\times 10^{-33}$, $y_{i}=1\times 10^{-41}$, $u_{i}=1.05 \times 10^{-29}$, ${\Varrho}_{i}=0.999877$. It is observed the two new scaling regimes during the radiation and dark matter era. To obtain this plot we have found the current values for the fractional energy densities of dark energy $\Omega_{de}^{(0)}=0.68$ and dark matter $\Omega_{m}^{(0)}=0.32$, at redshift $z=0$, according to Planck results \cite{Aghanim:2018eyx}. Also, during the scaling radiation epoch we have imposed the BBN constraint $\Omega_{de}^{(r)}<0.045$ \cite{Bean:2001wt}, and the constraint for the field energy density during the scaling matter epoch $\Omega_{de}^{(m)}<0.02$ ($95\%$ C.L.), at redshift $z\approx 50$, from CMB measurements \cite{Ade:2015rim}. } 
	\label{Figura1}
\end{figure}

\begin{figure}[htbp]
	\centering
		\includegraphics[width=0.45\textwidth]{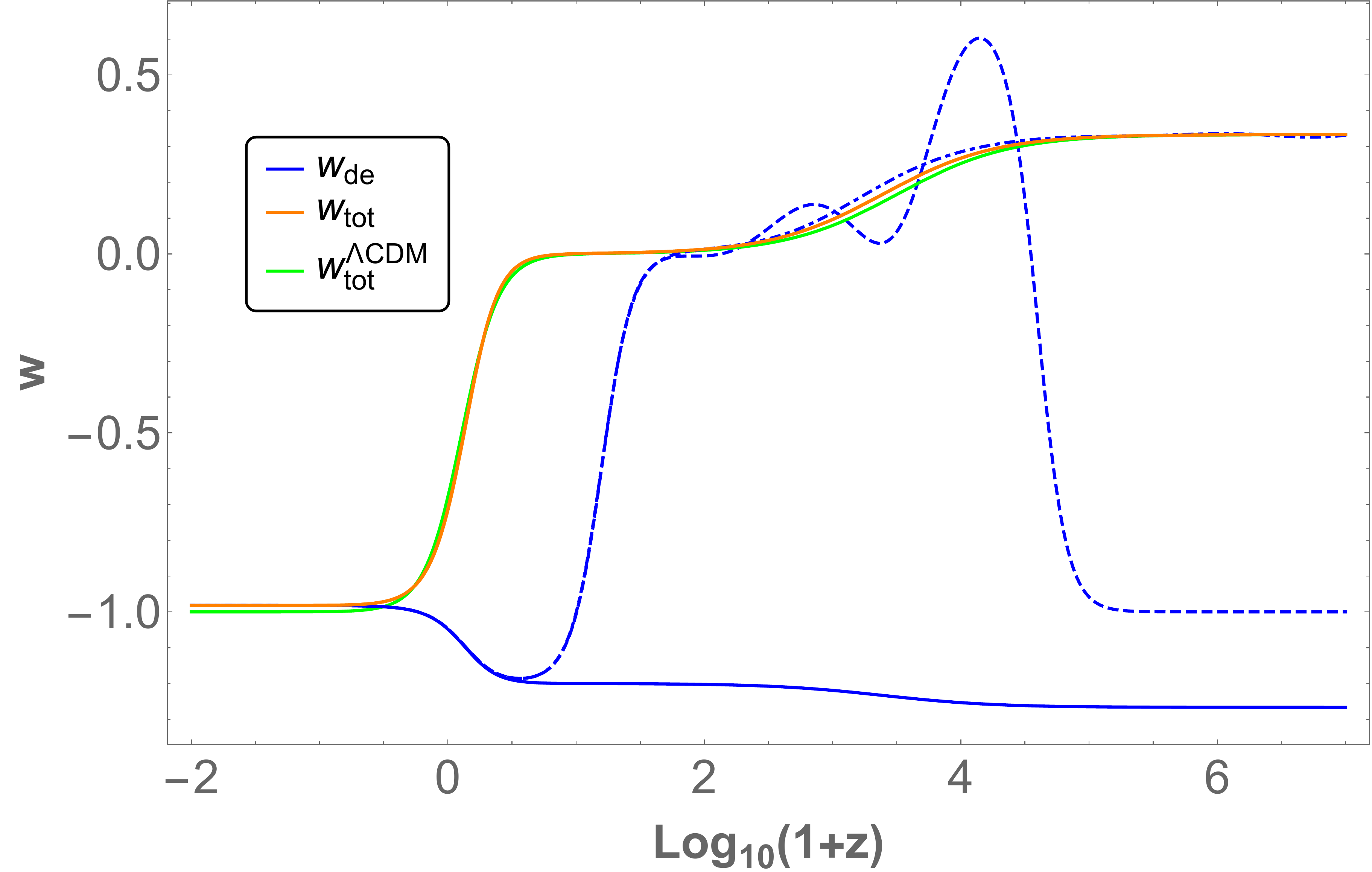}
	\caption{It is shown the behaviour of the total equation of state $w_{tot}$ (orange line), the equation of state of dark energy $w_{de}$ (blue line), and the total equation of state of $\Lambda$CDM model (green line) as functions of the redshift $z$, for the values of parameters $s=-1.20$, $\sigma=0.3$ and $\lambda=150$. Also, for solid, dashed, and dot-dashed blue lines we have the same initial conditions of FIG. \ref{Figura1}. It is observed the phantom value $w_{de}\approx -1.05$ at the current time $z=0$, which is consistent with the observational constraint $w_{de}^{(0)}=-1.028\pm 0.032$, from Planck data \cite{Aghanim:2018eyx}.}
	\label{Figura2}
\end{figure}

\begin{figure}[htbp]
	\centering
		\includegraphics[width=0.45\textwidth]{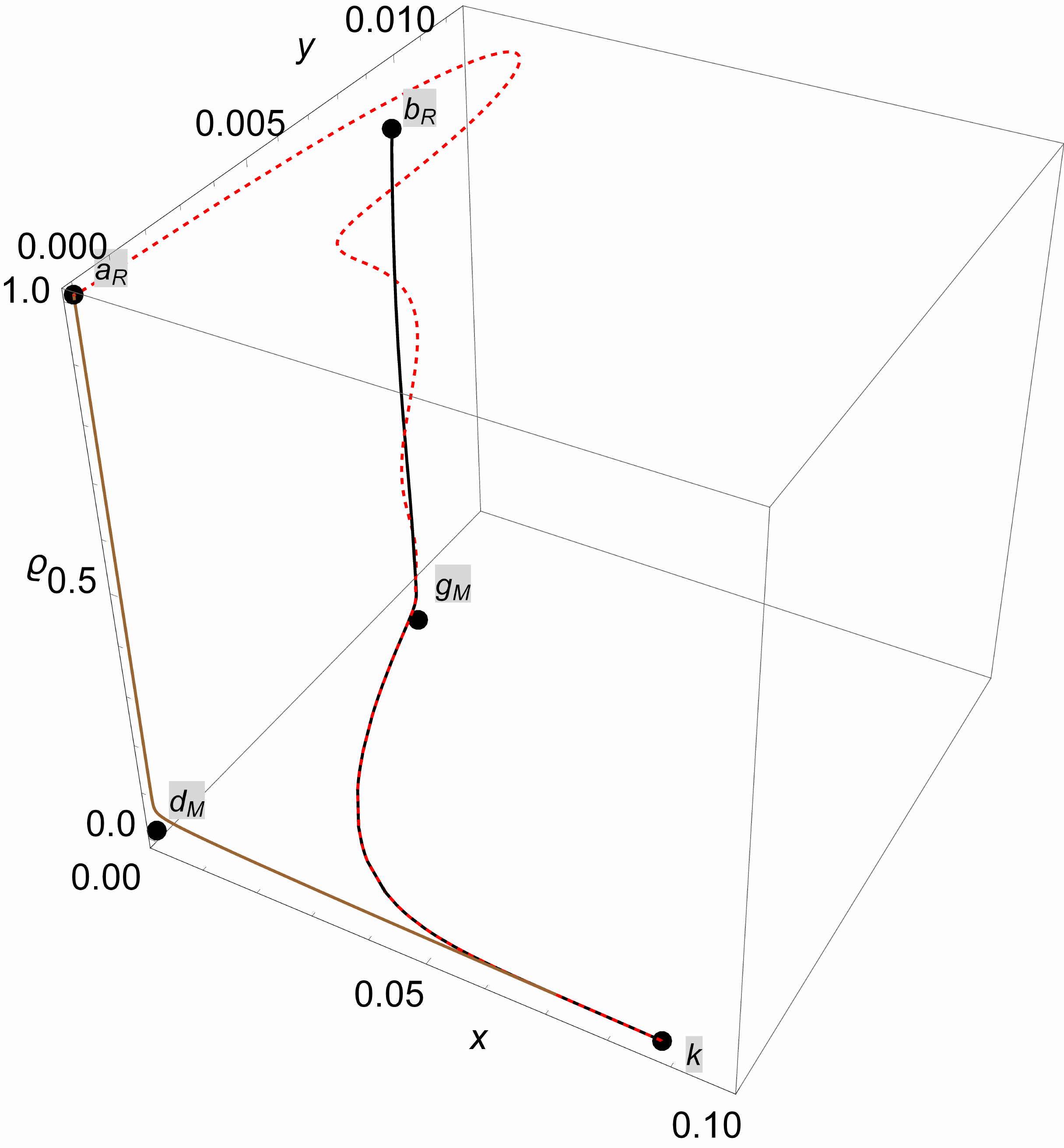}
	\caption{We plot the physical evolution curves in the phase space for the values of parameter $s=-1.20$, $\sigma=0.3$ and $\lambda=150$, and for the three different set of initial conditions $x_{i}=1\times 10^{-9}$, $y_{i}=4\times 10^{-7}$, $u_{i}=9.1\times 10^{-30}$, ${\Varrho}_{i}=0.999877$ (red dashed), $x_{i}=1\times 10^{-33}$, $y_{i}=1\times 10^{-41}$, $u_{i}=1.05 \times 10^{-29}$,${\Varrho}_{i}=0.999877$ (brown), and $x_{i}=0.0108$, $y_{i}=0.00765$, $u_{i}=9.3\times 10^{-30}$, ${\Varrho}_{i}=0.999788$ (black)  }
	\label{Figura3} 
\end{figure}

\begin{figure}[htbp]
	\centering
		\includegraphics[width=0.45\textwidth]{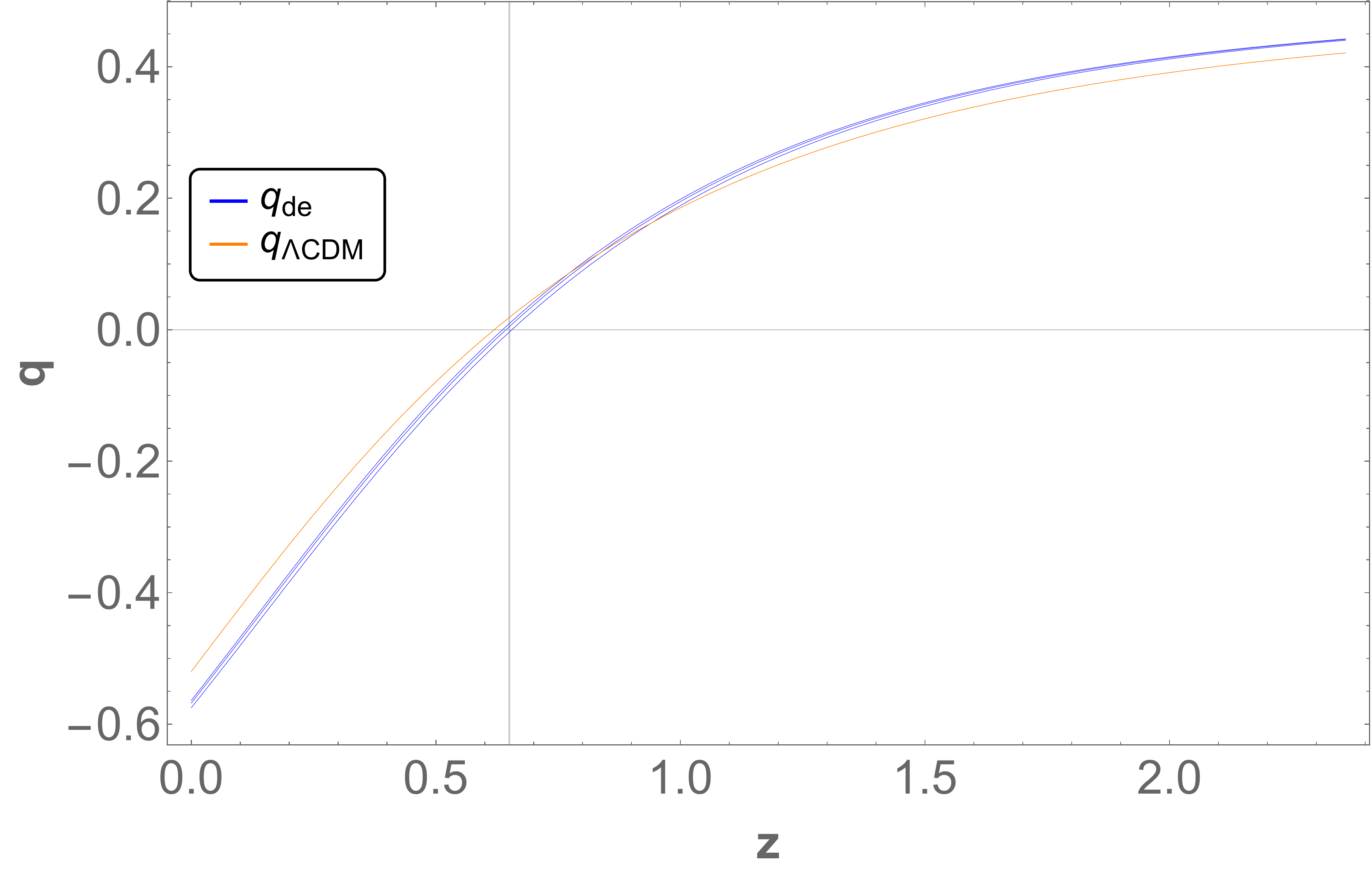}
	\caption{We depict the evolution of the deceleration parameter $q(z)$ as a function of the redshift $z$ for the values of parameters $s=-1.20$, $\sigma=0.3$ and $\lambda=150$, and the same initial conditions used in FIG. \ref{Figura1}. We have also depicted the corresponding curve for the deceleration parameter $q_{\Lambda CDM}(z)$ of the $\Lambda$CDM model. It is observed that in all the cases the cosmological deceleration-acceleration transition redshift happens at $z\approx 0.65 $, very close to $\Lambda$CDM value and consistent with current observational data \cite{Aghanim:2018eyx}.}
	\label{FIGq_k}
\end{figure}

\begin{figure}[htbp]
	\centering
		\includegraphics[width=0.45\textwidth]{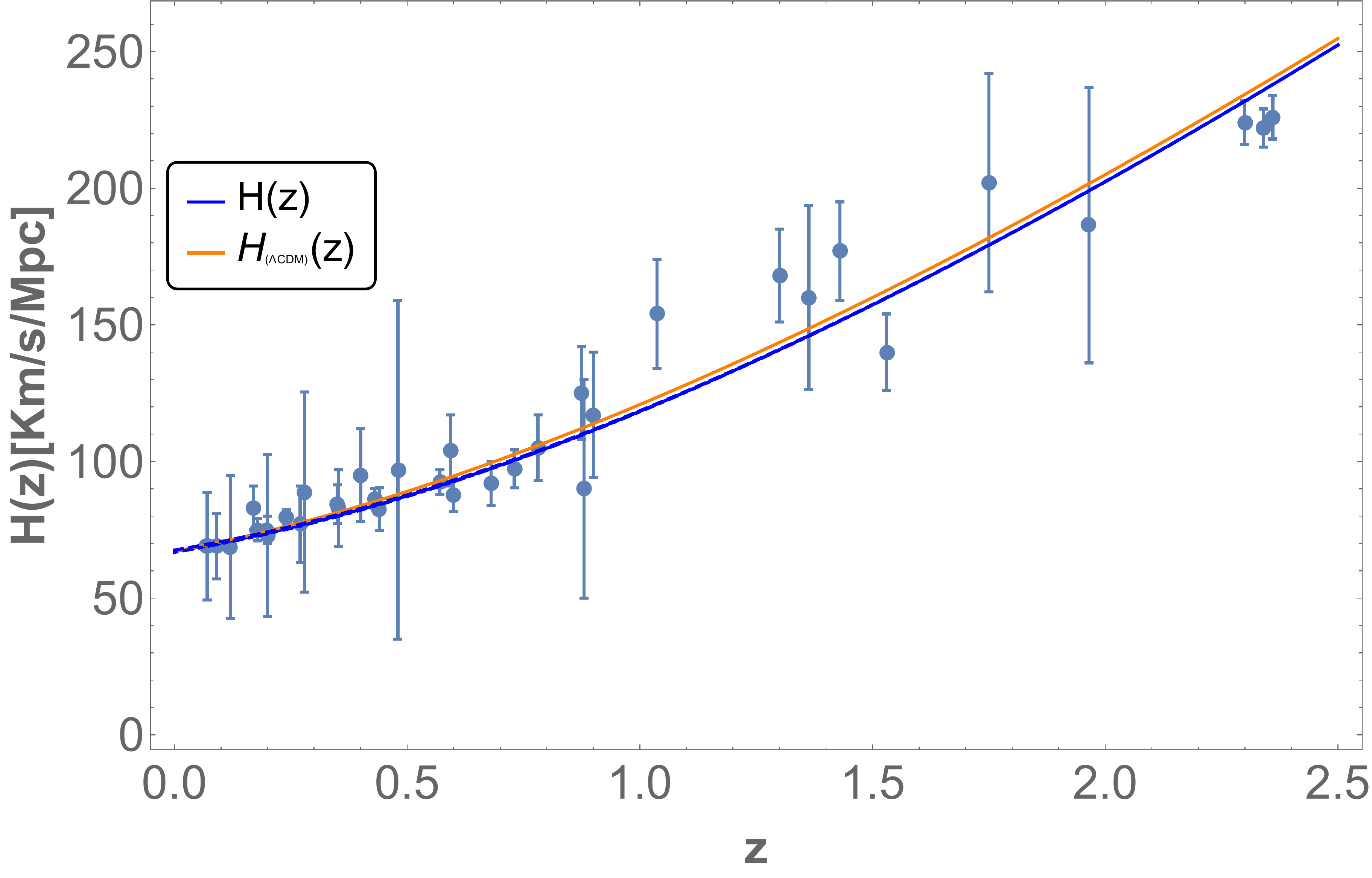}
	\caption{We show the evolution of the Hubble rate $H(z)$ as a function of the redshift $z$, for the values of parameters $s=-1.20$, $\sigma=0.3$ and $\lambda=150$, and the same initial conditions used in FIG. \ref{Figura1}, along with the evolution of the Hubble rate $H_{\Lambda CDM}(z)$ in the $\Lambda$CDM model and the Hubble data from Refs. \cite{Cao:2017gfv,Farooq:2013hq}. We have used the current value of the Hubble rate $H_{0}=67.4$ Km/(Mpc sec) from Planck 2018 \cite{Aghanim:2018eyx}.}
	\label{FIGH_k}
\end{figure}

\section{Numerical results}\label{Numerical}

We have found four final attractors which represent the dark energy-dominated epoch with cosmic acceleration, the points $h$, $k$, $j$ and $l$. The attractor $h$ is already present in the ordinary minimally coupled exponential quintessence model \cite{amendola2010dark}, but, $k$, $j$, and $l$ are new solutions which only arise in the non-minimal case and for $s\neq 0$.  Furthermore, it is found the scaling solutions $b_{R}$ and $c_{R}$ which are saddle points and represent the so-called scaling radiation era. The point $c_{R}$ is a new scaling solution which only exists in the non-minimal case and for $s\neq 0$. Also, we have found the scaling solutions $g_{M}$ and $i_{M}$ which are saddle points and they describe the so-called scaling dark matter era. Point $i_{M}$ only exists in the non-minimal case and $s\neq 0$.  

In FIGS. \ref{Figura1}, \ref{Figura2}, and \ref{Figura3}, we show the phase space trajectories $a_{R}\rightarrow d_{M}\rightarrow  k$, $a_{R}\rightarrow g_{M}\rightarrow  k$ and $b_{R}\rightarrow g_{M}\rightarrow  k$. In FIG. \ref{Figura1} we depict the behaviour of the energy densities of dark energy, dark matter and radiation, while in FIG. \ref{Figura2} we show the total equation of state and the equation of state of dark energy. The time of radiation-matter equality is around $z\approx 3387$, and the transition to the accelerated phase happens at $z\approx 0.65$, as it can be observed from the evolution curve of the deceleration parameter in FIG. \ref{FIGq_k}. This is very close to the $\Lambda$CDM value. Also, it is obtained the current values of the fractional energy density parameters of dark energy $\Omega_{de}^{(0)}\approx 0.68$ and dark matter $\Omega_{m}^{(0)}\approx 0.32$, with the equation of state of dark energy given by $w_{de}(z=0)\approx -1.048$ (dashed and dot-dashed blue lines), and $w_{de}(z=0)\approx -1.047$ (solid blue line), which is consistent with the observational constraint $w_{de}^{(0)}=-1.028\pm 0.032$, and with the other constraints for the cosmological parameters from Planck \cite{Aghanim:2018eyx}. Additionally, during the scaling radiation/matter regimes, for the evolution curves $a_{R}\rightarrow g_{M}\rightarrow  k$, and $b_{R}\rightarrow g_{M}\rightarrow  k$, we have applied the constraints on the fractional energy density parameters of dark energy, $\Omega_{de}^{(r)}$, $\Omega_{de}^{(m)}$, coming from the Physics of Big Bang Nucleosynthesis (BBN), $\Omega_{de}^{(r)}<0.045$ \cite{Bean:2001wt}, and CMB measurements from Planck, $\Omega_{de}^{(m)}<0.02$ ($95 \%$ CL), at redshift $z\approx 50$ \cite{Ade:2015rim}. For example, in FIG \ref{Figura1},  during the scaling radiation era $b_{R}$, we obtain $\Omega_{de}^{(r)}\approx 1.78\times 10^{-4}$ (solid blue line), and during the scaling matter era $g_{M}$, we find $\Omega_{de}^{(m)}\approx 1.35\times 10^{-4}$ (solid blue line), and  $\Omega_{de}^{(m)}\approx 1.37\times 10^{-4}$ (dashed blue line), at redshift $z=50$. Finally, in FIG. \ref{FIGH_k} we have depicted the evolution of the Hubble rate $H(z)$ for the present model using the above values of parameters and initial conditions, along with the evolution of Hubble rate $H_{\Lambda CDM}(z)$ of the $\Lambda$CDM model and the Hubble data from Refs. \cite{Cao:2017gfv,Farooq:2013hq}. It can be observed that the results obtained stayed very close to the $\Lambda$CDM results, and the present model passes the preliminary requirements to be considered as viable \cite{Aghanim:2018eyx}. 

\begin{figure}[htbp]
	\centering
		\includegraphics[width=0.45\textwidth]{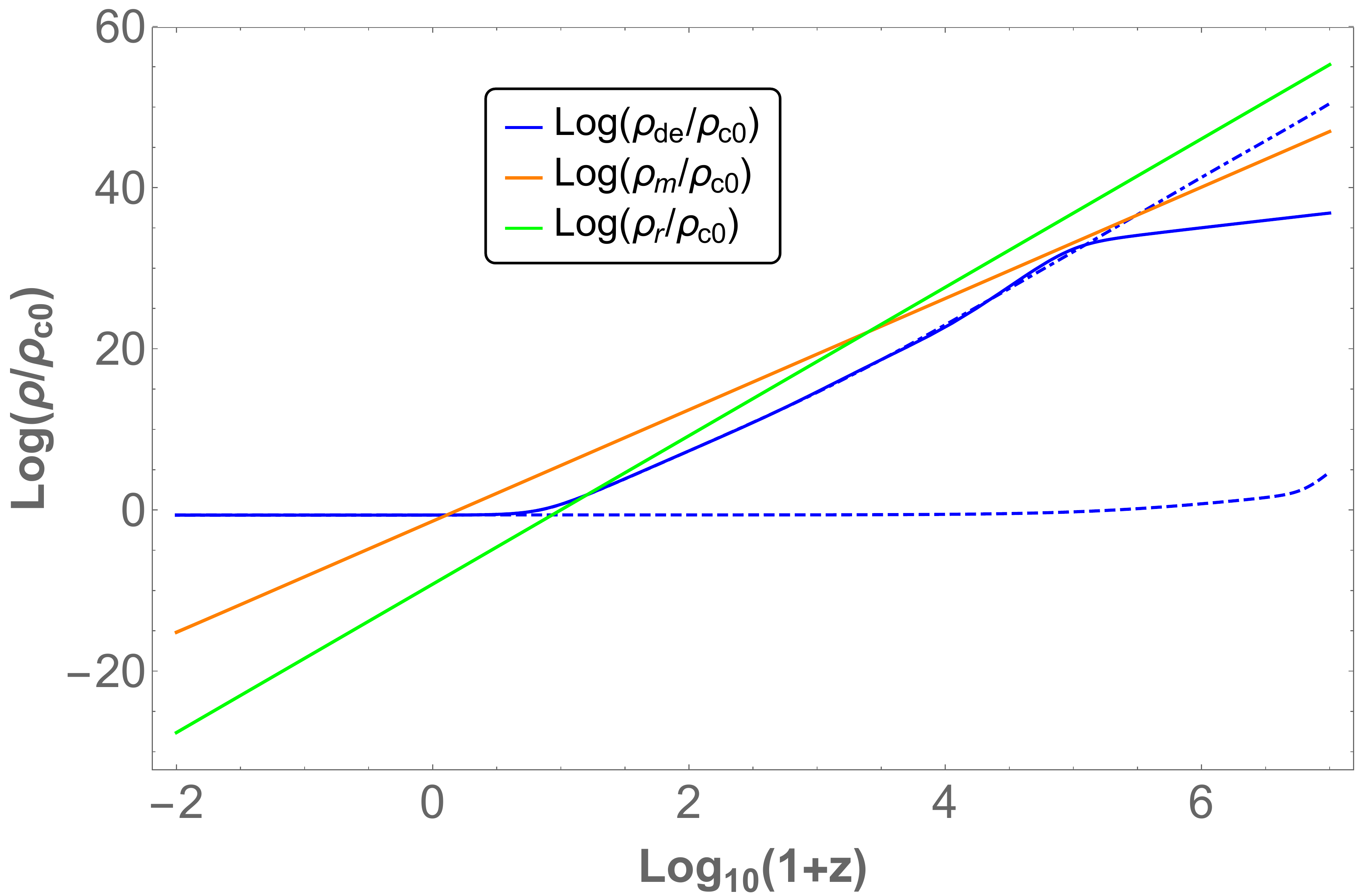}
	\caption{We depict the evolution of the energy density of dark energy $\rho_{de}$ (blue), dark matter (including baryons) $\rho_{m}$ (orange) and radiation $\rho_{r}$ (green) as functions of the redshift $z$, for $s=-0.8$, $\sigma=17$ and $\lambda=0.01$. The dashed line corresponds to the initial conditions $x_{i}=1\times 10^{-11}$, $y_{i}=7.3\times 10^{-13}$, $u_{i}=1\times 10^{-8}$, ${\Varrho}_{i}=0.999875$, solid blue line to $x_{i}=1\times 10^{-8}$, $y_{i}=7.3\times 10^{-13}$, $u_{i}=1\times 10^{-23}$, ${\Varrho}_{i}=0.999875$, and dot-dashed blue line to $x_{i}=0.0768467$, $y_{i}=7.3\times 10^{-13}$, $u_{i}=0.002$, ${\Varrho}_{i}=0.995916$. It is observed the two new scaling regimes during the radiation and dark matter era. To obtain this plot we have found the current values for the fractional energy densities of dark energy $\Omega_{de}^{(0)}=0.68$ and dark matter $\Omega_{m}^{(0)}=0.32$, at redshift $z=0$, according to Planck results \cite{Aghanim:2018eyx}. Also, during the scaling radiation epoch we have imposed the BBN constraint $\Omega_{de}^{(r)}<0.045$ \cite{Bean:2001wt}, and the constraint for the field energy density during the scaling matter epoch $\Omega_{de}^{(m)}<0.02$ ($95\%$ C.L.), at redshift $z\approx 50$, from CMB measurements \cite{Ade:2015rim}. }
	\label{Figura4}
\end{figure}

\begin{figure}[htbp]
	\centering
		\includegraphics[width=0.45\textwidth]{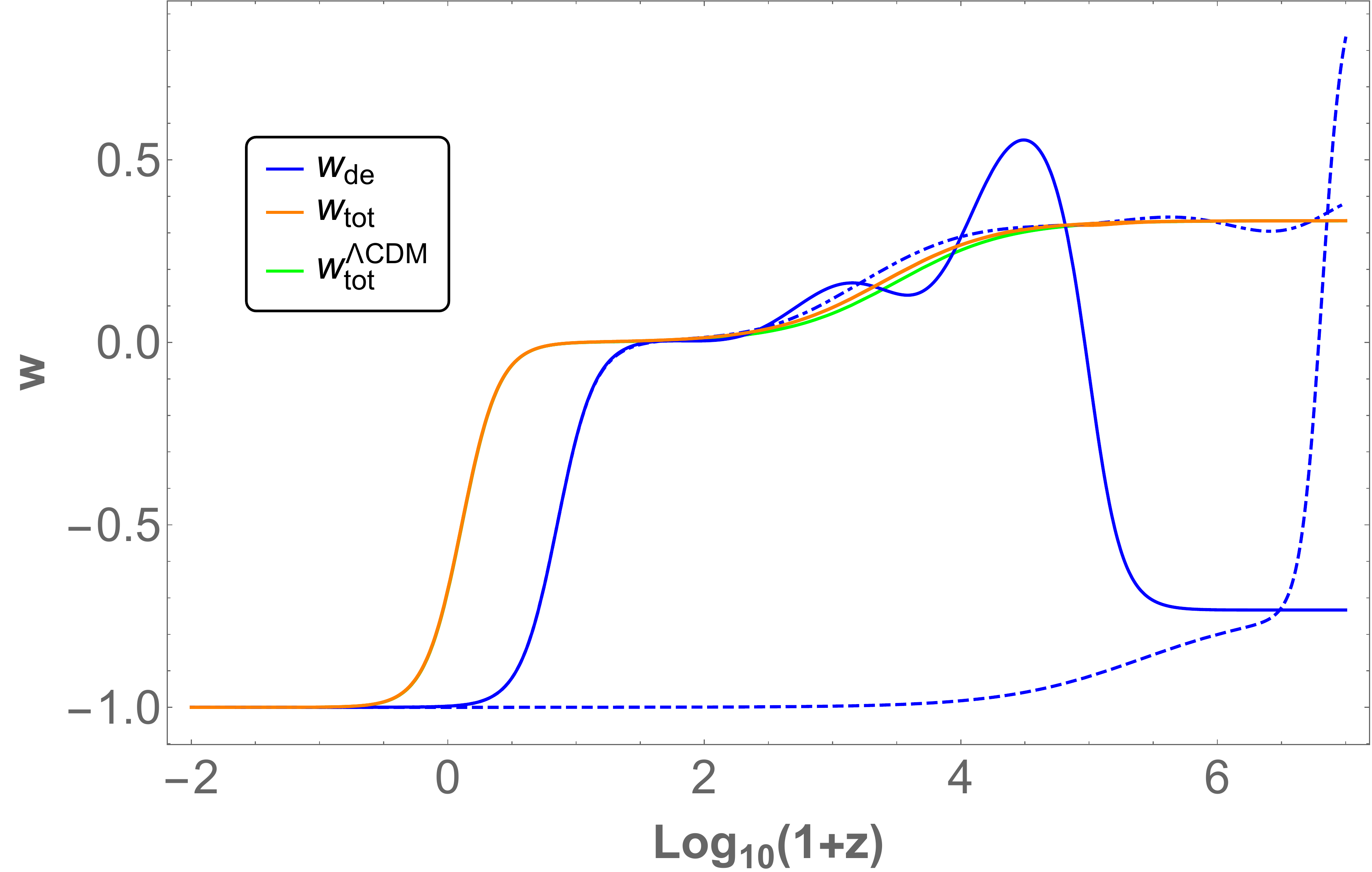}
	\caption{It is shown the behaviour of the total equation of state $w_{tot}$ (orange line), the equation of state of dark energy $w_{de}$ (blue line), and the total equation of state of $\Lambda$CDM model (green line) as functions of the redshift $z$ for $s=-0.8$, $\sigma=17$ and $\lambda=0.01$. Also, for solid, dashed, and dot-dashed blue lines we have the same initial conditions of FIG. \ref{Figura4}. It is obtained the value $w_{de}\approx -0.997$ at the current time $z=0$, which is consistent with the observational constraint $w_{de}^{(0)}=-1.028\pm 0.032$, from Planck data \cite{Aghanim:2018eyx}. }
	\label{Figura5}
\end{figure}

\begin{figure}[htbp]
	\centering
		\includegraphics[width=0.45\textwidth]{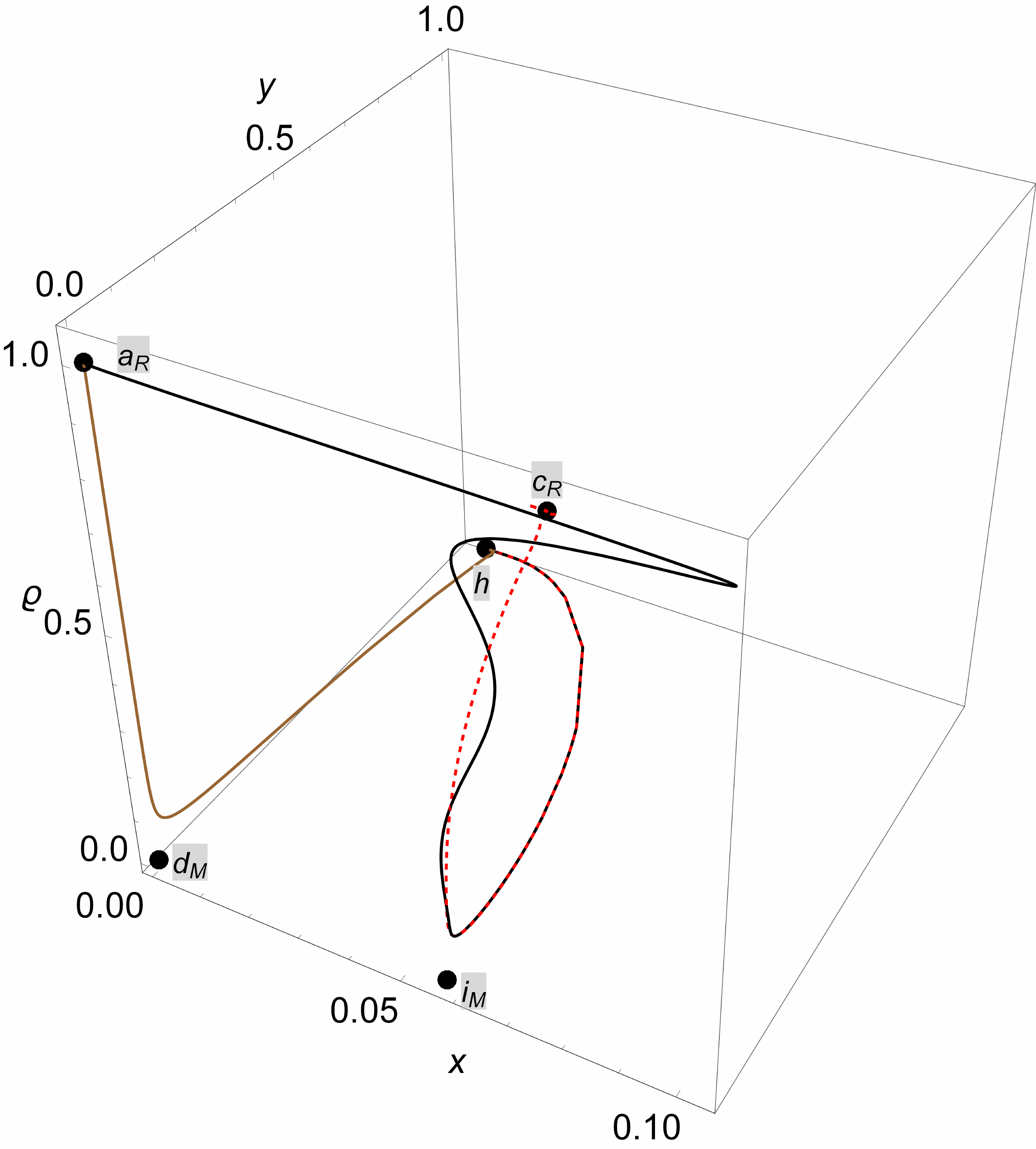}
	\caption{We plot the physical evolution curves in the phase space for the values of parameter $s=-0.8$, $\sigma=17$ and $\lambda=0.01$, and for the three different set of initial conditions $x_{i}=1\times 10^{-11}$, $y_{i}=7.3\times 10^{-13}$, $u_{i}=1\times 10^{-8}$, ${\Varrho}_{i}=0.999875$ (brown), $x_{i}=1\times 10^{-8}$, $y_{i}=7.3\times 10^{-13}$, $u_{i}=1\times 10^{-23}$, ${\Varrho}_{i}=0.999875$ (black), and  $x_{i}=0.0768467$, $y_{i}=7.3\times 10^{-13}$, $u_{i}=0.002$, ${\Varrho}_{i}=0.995916$ (red dashed).}
	\label{Figura6}
\end{figure}

\begin{figure}[htbp]
	\centering
		\includegraphics[width=0.45\textwidth]{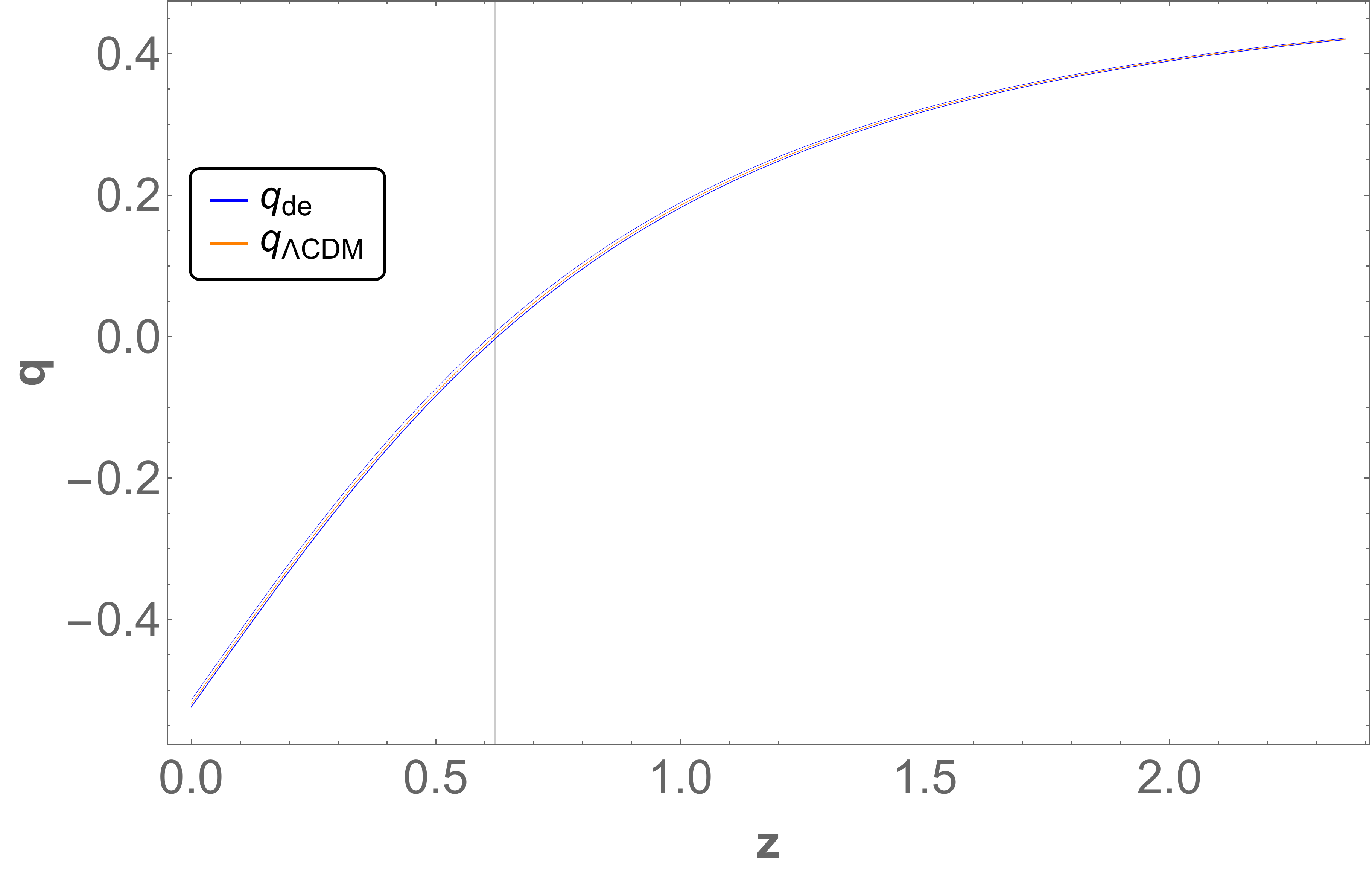}
	\caption{We depict the evolution of the deceleration parameter $q(z)$ as a function of the redshift $z$ for the values of parameters $s=-0.8$, $\sigma=17$ and $\lambda=0.01$, and the same initial conditions used in FIG. \ref{Figura4}. It is also shown the corresponding curve for the deceleration parameter $q_{\Lambda CDM}(z)$ of the $\Lambda$CDM model. It is observed that in all the cases the cosmological deceleration-acceleration transition redshift happens at $z\approx 0.62 $, very close to the $\Lambda$CDM value and then it is consistent with current observational data \cite{Aghanim:2018eyx}.}
	\label{FIGq_h}
\end{figure}

\begin{figure}[htbp]
	\centering
		\includegraphics[width=0.45\textwidth]{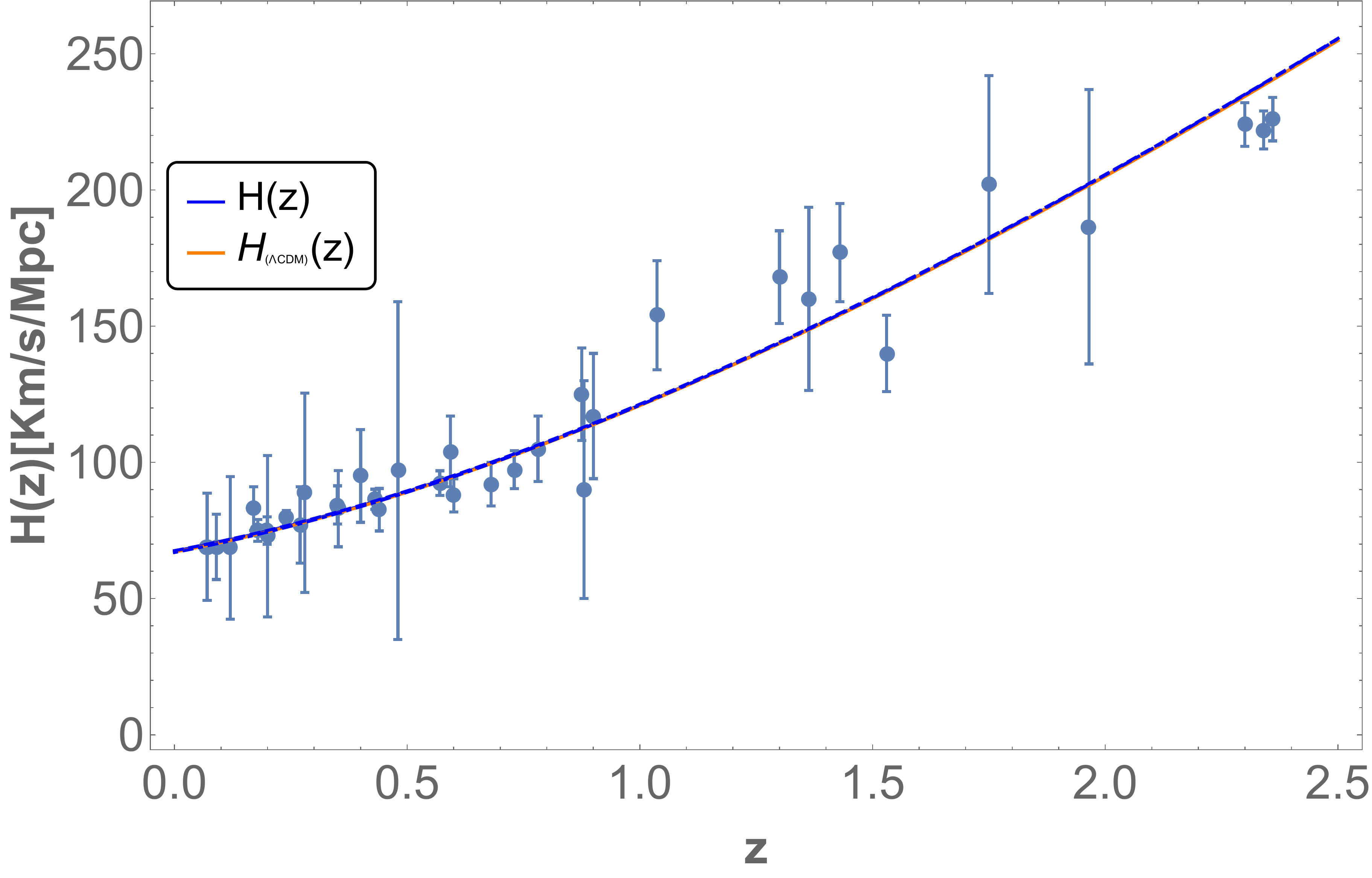}
	\caption{We show the evolution of the Hubble rate $H(z)$ as a function of the redshift $z$, for the values of parameters $s=-0.8$, $\sigma=17$ and $\lambda=0.01$, and the same initial conditions used in FIG. \ref{Figura1}, along with the evolution of the Hubble rate $H_{\Lambda CDM}(z)$ of the $\Lambda$CDM model and the Hubble data from Refs. \cite{Cao:2017gfv,Farooq:2013hq}. We have used the current value of the Hubble rate $H_{0}=67.4$ Km/(Mpc sec) from Planck 2018 \cite{Aghanim:2018eyx}.}
	\label{FIGH_h}
\end{figure}

Similarly, in FIGS. \ref{Figura4}, \ref{Figura5}, and \ref{Figura6}, we show the evolution curves in the phase space for the transitions $a_{R}\rightarrow d_{M}\rightarrow  h$, $a_{R}\rightarrow i_{M}\rightarrow  h$ and $c_{R}\rightarrow i_{M}\rightarrow  h$. In FIGS. \ref{Figura4}, and \ref{Figura5}, we plot the evolution of the energy densities of the matter components and the total equation of state and the equation of state of dark energy, respectively. As before, the redshift of radiation-matter equality is around $z\approx 3387$, and the transition to the accelerated phase at $z\approx 0.62$ (See FIG. \ref{FIGq_h}), very close to the $\Lambda$CDM value. Also, these evolution trajectories can adjust the current values of the fractional energy density parameters of dark energy $\Omega_{de}^{(0)}\approx 0.68$ and dark matter $\Omega_{m}^{(0)}\approx 0.32$, and the equation of state of dark energy now takes the value $w_{de}(z=0)\approx -0.9968$ (solid and dot-dashed blue lines), and $w_{de}(z=0)\approx -1$ (dashed blue line), which is again consistent with the observational constraints from Planck \cite{Aghanim:2018eyx}. Likewise, during the scaling radiation era represented by the critical point $c_{R}$, we have 
$\Omega_{de}^{(r)}\approx 8.1\times 10^{-3}$ (dot-dashed blue line), which is consistent with the BBN constraint \cite{Bean:2001wt}, and for the scaling matter era represented by point $i_{M}$ we get $\Omega_{de}^{(m)}\approx 5.89\times 10^{-3}$ (solid blue line), and $\Omega_{de}^{(m)}\approx 5.85\times 10^{-3}$ (dot-dashed blue line), at $z=50$, which is also consistent with CMB measurements \cite{Ade:2015rim}. Furthermore, in FIG. \ref{FIGq_h} we also show the evolution of the Hubble rate $H(z)$, along with the evolution of Hubble rate $H_{\Lambda CDM}(z)$ of the $\Lambda$CDM model and the Hubble data from Refs. \cite{Cao:2017gfv,Farooq:2013hq}. Then, the results that we have found here are very close to the $\Lambda$CDM results, and so, the model satisfies the preliminary requirements to be considered as viable \cite{Aghanim:2018eyx}.

\begin{table*}[b]
 \centering
 \caption{Properties of the critical points.}
\begin{center}
{
\begin{tabular}{c c c c}\hline\hline
Name & Existence & Stability & Acceleration \\\hline
$a_{R}$ & $\forall~ s, \lambda, \sigma$ & unstable $\forall~ ~ s, \lambda, \sigma $ & never \\
$b_{R}$ &  $\abs{\lambda}>2$ & unstable $\forall~ ~ s, \lambda, \sigma $ & never  \\
$c_{R}$ & $\sigma \neq 0$  & unstable $\forall~ ~ s, \lambda, \sigma $ & never   \\
 & $\land(-\frac{1}{8} \sqrt{12 \sigma ^2+1}-\frac{1}{8}<s<-\frac{1}{4})$  & unstable $\forall~ ~ s, \lambda, \sigma $ & never   \\
 & $\lor(0<s<\frac{1}{8} \sqrt{12 \sigma ^2+1}-\frac{1}{8})$  & unstable $\forall~ ~ s, \lambda, \sigma $ & never   \\
$d_{M}$ &  $\forall~ s, \lambda, \sigma$ & unstable $\forall~ ~ s, \lambda, \sigma $ & never  \\
$e$ & $\forall~ s, \lambda, \sigma$ & unstable $\forall~ ~ s, \lambda, \sigma $ & never  \\
$f$ & $\forall~ s, \lambda, \sigma$ & unstable $\forall~ ~ s, \lambda, \sigma $ & never  \\
$g_{M}$ &  $\abs{\lambda}>\sqrt{3}$ & unstable for $s\in \mathbb{R}$ & never  \\
&  & $\land \left(\lambda <-\sqrt{3}\land \sigma >-\lambda s\right)$ &   \\
&  & $\lor \left(\lambda >\sqrt{3}\land \sigma <-\lambda s\right)$ &   \\
$h$ &  $\abs{\lambda}<\sqrt{6}$  & $s\in \mathbb{R}$ & $\abs{\lambda}<\sqrt{2}$ \\
 &  & $\land \left(-\sqrt{2}<\lambda <0\land \sigma <-\lambda s\right)$ &  \\
 &  & $\lor \left(0<\lambda <\sqrt{2}\land \sigma >-\lambda s\right)$ &  \\
$i_{M}$ &  $s>0 \land \abs{\sigma}> \sqrt{\frac{3s(3 s+1)}{2}}$ & unstable for & never\\
 &  $s<-\frac{1}{3} \land \abs{\sigma}> \sqrt{\frac{3s(3 s+1)}{2}}$  & $s>0 \land \abs{\sigma}>\sqrt{\frac{3 s (s+1) (2 s+1)}{2}}$ & \\
&   & $-\frac{1}{6} \sqrt{8 \sigma ^2+1}-\frac{1}{6}<s<-\frac{1}{3}$ & \\
&   & $\land (\sigma >0\land \lambda <-\frac{\sigma }{s})$ & \\
&    & $\lor (\sigma <0\land \lambda >-\frac{\sigma }{s})$ & \\
$j$ &  $s>0 \land \abs{\sigma} \leq \sqrt{\frac{3 s (2 s+1)^2}{2(s+1)}}$   & $-1<s<0$  & $\sigma \in \mathbb{R}\land -1<s<0$ \\
& $s<-1 \land \abs{\sigma} \leq \sqrt{\frac{3 s (2 s+1)^2}{2(s+1)}}$   & $ \land (\lambda \leq 0\land (\sigma <-\lambda s\lor \sigma >0))$ &  \\
&  $-1\leq s<0$ and $\sigma \in \mathbb{R}$   &  $\lor (\lambda >0\land (\sigma <0\lor \sigma >-\lambda s))$  &  \\
$k$ &  $s>0 \land \abs{\sigma} \leq \sqrt{\frac{3 s (2 s+1)^2}{2(s+1)}}$  &  $s>\sqrt{\frac{3 \sigma ^2}{10}+\frac{1}{25}}-\frac{1}{5}$ & $\abs{\sigma}<\sqrt{\frac{2s (5 s+2)}{3}}$ \\
&  $s<-1 \land \abs{\sigma} \leq \sqrt{\frac{3 s (2 s+1)^2}{2(s+1)}}$   & $\land (\sigma <0\land \lambda >-\frac{\sigma }{s})$ & $\land~s>0$ \\
& $-1\leq s<0$ and $\sigma \in \mathbb{R}$  &  $\lor (\sigma >0 \land \lambda <-\frac{\sigma }{s})$ & $\lor \left(-1<s<-\frac{2}{5}\lor s<-1\right)$  \\
&  &  $s<-\sqrt{\frac{3 \sigma ^2}{10}+\frac{1}{25}}-\frac{1}{5}$ & \\
&  &  $ \land (\sigma \leq -\sqrt{2}\land \lambda <-\frac{\sigma }{s})$ &   \\
&   &  $\lor (\sigma >\sqrt{2}\land \lambda >-\frac{\sigma }{s})$ &  \\
&   &  $s<-1$    &            \\
&   &  $\land (-\sqrt{2}<\sigma <0\land \lambda <-\frac{\sigma }{s})$ & \\
&   &  $\lor (0<\sigma \leq \sqrt{2}\land \lambda >-\frac{\sigma }{s})$ &   \\
&   &  $-1<s<-\sqrt{\frac{3 \sigma ^2}{10}+\frac{1}{25}}-\frac{1}{5}$  &     \\
&   &   $ \land (-\sqrt{2}<\sigma <0\land \lambda <-\frac{\sigma }{s})$  &    \\
&   &   $\lor (0<\sigma <\sqrt{2}\land \lambda >-\frac{\sigma }{s})$  &   \\
$l$ &  $\sigma <0\land s>0\land \lambda <-\frac{\sigma }{2 s+1}$  & $\lambda >0\land s>0 ~\land$ & always \\
&  $\sigma >0\land s>0\land \lambda >-\frac{\sigma }{2 s+1}$  & $ 0<\sigma \leq \frac{-4 \lambda ^2 s+\lambda  \sqrt{\frac{9}{\lambda ^2}+8 s \left(2 \left(\lambda ^2+6\right) s+9\right)}-3}{8 \lambda }$ &  \\
&  $\sigma <0\land s<-\frac{1}{2}\land \lambda >-\frac{\sigma }{2 s+1}$  &  &  \\
&  $\sigma >0\land s<-\frac{1}{2}\land \lambda <-\frac{\sigma }{2 s+1}$  &  &  \\
&  $\sigma <0\land -\frac{1}{2}<s<0\land \lambda <-\frac{\sigma }{2 s+1}$  &  &  \\
&  $\sigma >0\land -\frac{1}{2}<s<0\land \lambda >-\frac{\sigma }{2 s+1}$  &  &  \\
\\ \hline\hline
\end{tabular}}
\end{center}
\label{table3}
\end{table*}

\section{Concluding remarks}\label{Conclu}

In the present work we have investigated the cosmological dynamics of dark energy in the context of scalar-torsion $f(T,\phi)$  gravity, where $f(T,\phi)$ is a function of the torsion scalar $T$, associated with the Weitzenb\"{o}ck connection in the context of modified teleparallel gravity, and the scalar field $\phi$. Particularly, we have studied the class of theories with Lagrangian density $f(T,\phi)=-T/2 \kappa^2-F(\phi) G(T)-V(\phi)$, with $F(\phi)\sim e^{-\sigma \kappa\phi}$, $V(\phi)\sim e^{-\lambda \kappa\phi}$ and $G(T)\sim T^{1+s}$,  plus the canonical kinetic term for the scalar field. The exponential scalar potential is the usual one studied in the context of dark energy which admits scaling solutions, while the exponential coupling function of the scalar field is the simplest natural choice to be assumed with this potential \cite{Amendola:1999qq}. Furthermore, the function $G(T)$ generalises the typical choice of a linear function of $T$ for the non-minimal coupling to gravity \cite{Geng:2011aj,Otalora:2013tba} to a non-linear case. It has been shown in Ref. \cite{Gonzalez-Espinoza:2020azh} that in order to generate primordial fluctuations during inflation from $f(T,\phi)$ gravity, non-linear terms in torsion scalar need to be considered to construct the coupling function, when the non-minimal coupling to gravity is switch on. These non-linear scalar-torsion coupling terms can also be seen as a torsion-based analogue of non-linear matter-gravity couplings in extended $f(R)$ gravity theories \cite{Nojiri:2004bi,Allemandi:2005qs,Nojiri:2006ri}, but neglecting the coupling to the kinetic term of the scalar field. Including the kinetic term in the non-minimal coupling function could deviate the squared tensor propagation speed from $1$ \cite{Gonzalez-Espinoza:2019ajd}, which is something undesirable \cite{Baker:2017hug,Sakstein:2017xjx}. 

For the FLRW background, and in the presence of radiation and cold dark matter, we have defined the effective energy and pressure densities of dark energy. We have obtained the autonomous system associated with the set of cosmological equations and then we have performed the dynamical analysis in the phase space by getting the critical points, their cosmological properties and stability conditions. The critical points are presented in the Table \ref{table1}, while the expressions for the cosmological parameters for each critical point are shown in Table \ref{table2}. Finally, the conditions of existence, stability and acceleration are shown in Table \ref{table3}. From these results we have shown that the dark energy model at hand is cosmologically viable since the thermal history of the Universe is successfully reproduced. For instance, the corresponding physical evolution curves in the phase space are depicted in FIG. \ref{Figura3} and FIG. \ref{Figura6}. Let us remember that for any dark energy model to be viable it is required that the dark energy remains subdominant during the radiation and matter dominating eras, emerging only at late-times to produce the current accelerated expansion of the Universe \cite{amendola2010dark}. From the point of view of the dynamical systems theory, it is required the existence of unstable critical points with decelerated expansion representing the radiation and matter dominated periods, and stable critical points (attractors) with accelerated expansion to describe the dark energy phase \cite{Copeland:2006wr,amendola2010dark}.

The modified gravity extension of the teleparallel equivalent of general relativity, or teleparallel gravity \cite{Einstein,TranslationEinstein,Early-papers1,Early-papers2,Early-papers3,Early-papers4,Early-papers5,Early-papers6,Aldrovandi-Pereira-book,JGPereira2,AndradeGuillenPereira-00,Arcos:2005ec,Pereira:2019woq} for short, namely, $f(T)$ gravity \cite{Bengochea:2008gz,Linder:2010py} has been proposed as a good alternative to the curvature-based modified gravity theories like $f(R)$ gravity \cite{Clifton:2011jh,Capozziello:2011et,DeFelice:2010aj,Nojiri:2010wj,Nojiri:2006ri}, to explain the current accelerated expansion of the Universe, and its cosmological dynamics has been studied in detail in Refs. \cite{Wu:2010xk,Zhang:2011qp,Jamil:2012nma,Ganiou:2018dta}. However, as it has been shown in Refs. \cite{Wu:2010xk,Cai:2015emx,Gonzalez-Espinoza:2020azh}, only a marginally stable de Sitter solution, and not a de Sitter attractor, is found when analysing the dynamics of $f(T)$ gravity. Let us remember that an attractor condition of a fixed point of any autonomous system is a special feature of such solution which allows that sooner or later the system reaches this critical point for generic initial conditions \cite{Copeland:2006wr,Bahamonde:2017ize}. This is a highly desired property of any dark energy solution, either a de Sitter, or scaling solution, in order to explain why dark energy has to come to dominate just at late-times without a fine-tuning of the initial conditions \cite{Copeland:2006wr,amendola2010dark}.

On the other hand, it is well known that scalar fields are commonly present in particle physics, string theory, and cosmology,  \cite{Fujii:2003pa, Faraoni:2004pi}. Furthermore, the non-minimal coupling terms between the scalar field and curvature can arise from a variety of model-building efforts in theoretical physics (see also Refs. \cite{Fujii:2003pa, Faraoni:2004pi}), and they are required as counterterms in the process of quantisation of the scalar field in a curved space-time \cite{Birrell:1982ix}. In the context of cosmology, it is well known that the introduction of a non-minimally coupled scalar field is very healthy and it can give place to the existence of new dark energy attractors, and scaling solutions that play an important role in alleviating the coincidence and energy scale problems of dark energy \cite{Albuquerque:2018ymr,Ohashi:2009xw}. So, although from a general theoretical point of view the increase in the number of degrees of freedom of the system to explain the corresponding phenomenon could lead to a weaker motivation for that model, in the present case, the addition of one more degree of freedom in scalar-torsion $f(T,\phi)$ gravity is very well motivated, and it is very closely related to similar constructions based on curvature like scalar-tensor gravity \cite{Fujii:2003pa, Faraoni:2004pi}, and generalised scalar-tensor $f(R,\phi)$ gravity theories \cite{Faraoni:2004pi,Tsujikawa:2008uc,Alimohammadi:2009yt}.

Some interesting features have been found that make the model appealing as a viable dark energy candidate. For the non-linear coupling function $G(T)\sim T^{1+s}$, with $s\neq 0$, we have obtained new attractors solutions describing the dark energy-dominated era and new scaling solutions representing the scaling radiation/matter eras. Since we have found that the latter are saddle points, we got scaling regimes during the radiation and cold dark matter epochs followed by the dark energy attractor with accelerated expansion. In FIGS. \ref{Figura1} and \ref{Figura4}, as well as in FIGS. \ref{Figura3}, and \ref{Figura6}, we have numerically confirmed that the dynamics of the model can allow the two scaling regimes previous to the dark energy-dominated epoch, satisfying the cosmological constraints for the early-time dark energy density from BBN \cite{Bean:2001wt} and CMB bounds \cite{Aghanim:2018eyx}. Therefore, the final attractor can be either a de Sitter solution ($w_{de}=-1$), or a dark energy-dominated solution with $\Omega_{de}=1$, and equation of state with quintessence-like, phantom-like behaviour or experiencing the phantom-divide crossing as illustrated in FIGS. \ref{Figura2} and \ref{Figura5}. We have also determined some ranges for the parameters of the scaling solutions where there is also stability of them, but in this case, these solutions cannot explain the current accelerated expansion. On other other hand, the phantom-divide crossing during the cosmological evolution indicates a distinctive feature of our model as compared with the minimally coupled scalar field in GR, and it has been inherited from the linear case $G(T)\sim T$ for $s=0$ \cite{Geng:2011aj}. 

These consequences, namely, the existence of new dark energy attractors and new scaling solutions, are originated by the addition of the scalar field $\phi$ non-minimally coupled to the non-linear function $G(T)\sim T^{1+s}$. Furthermore, these solutions are not present neither in pure $f(T)$ gravity \cite{Wu:2010xk,Zhang:2011qp,Jamil:2012nma,Ganiou:2018dta} nor in non-minimally coupled scalar field models with a linear coupling to torsion ($s=0$), the so-called teleparallel dark energy model \cite{Geng:2011aj,Wei:2011yr,Xu:2012jf}. In fact, the new dark energy attractors are the fixed points $j$ and $k$, that are field dominated solutions, and the fixed point $l$ which is a de Sitter solution. The new scaling solutions are the fixed points $c_{R}$ and $i_{M}$. All these critical points along with the remaining points have been summarised in Table \ref{table1}. We have parametrised the non-linearity of the coupling between the scalar field and torsion through the parameter $s$, being that the linear case (teleparallel dark energy) is recovered for $G(T)\sim T$, that is to say $s=0$. For $s=0$, it is seen that $c_{R}$ becomes equal to the standard radiation era represented by fixed point $a_{R}$, while $i_{M}$ converts to the standard matter era $d_{M}$. So, the scaling solutions $c_{R}$ and $i_{M}$ only exist in the case of a non-linear coupling to torsion such that $G(T)\sim T^{1+s}$, with $s\neq 0$. Similarly, points $j$, $k$, and $l$ only exist for the non-linear case  $G(T)\sim T^{1+s}$, with $s\neq 0$, once that for $s=0$ these points are lost. In particular, for points $j$ and $k$, with $s=0$, we fall into another different coupling-dominated solution. Also, for $s=0$, point $l$ yields another de Sitter solution which exists only for $G(T)\sim T$.

In FIGS. \ref{FIGH_k}, and \ref{FIGH_h}, we have also depicted the evolution of the Hubble rate $H(z)$, along with the Hubble data from Refs. \cite{Cao:2017gfv,Farooq:2013hq}, corroborating that the results obtained here are very close to the $\Lambda$CDM results for $H(z)$, and so the present model satisfies the preliminary requirements to be considered as viable \cite{Aghanim:2018eyx}. Even more, it is important to highlight that due to the existence of new scaling solutions that naturally incorporate the early dark energy, there is an additional phenomenological interest in the present model that does not happen in the case of the $\Lambda$CDM model. The scaling solutions provide a mechanism to alleviate the energy scale problem of the $\Lambda$CDM model related to the large energy gap between the critical energy density of the Universe today and the typical energy scales of particle physics \cite{Albuquerque:2018ymr,Ohashi:2009xw}. This is due to that during a scaling radiation$/$matter regime the field energy density is not necessarily negligible compared to the energy density of the background fluid at early times (see FIGS. \ref{Figura1} and \ref{Figura4}). Also, a model that predicts a dark energy component during the early universe is strongly constrained, and it also may lead to new imprints in the early-time physics that can allow to distinguish it from the other alternatives to explain dark energy \cite{Koivisto:2008xf,Doran:2006kp}.   

The principal aim of our work has been to model dark energy from the point of view of dynamical systems in the context of scalar-torsion $f(T,\phi)$ gravity where $T$ is the torsion scalar of teleparallel gravity and $\phi$ is a canonical scalar field. Thus, after calculating the modified Friedmann equations for the model at hand, we have written these equations in the standard form by identifying the effective energy and pressure densities of dark energy, Eqs. \eqref{rho_de} and \eqref{p_de}, respectively. So, from these expressions it is recognized that dark energy could be originated from both contributions, namely, the energy density of the scalar field and the generalised non-minimal coupling to torsion. In fact, by analysing the critical points of the associated autonomous system we have found new dark energy solutions that only exist for the case of the non-linear coupling to torsion, $G(T)\sim T^{1+s}$ with $s\neq 0$. Since we have verified that these new dark energy solutions are attractors points the universe will definitely reach these fixed points for a wide range of initial conditions. Furthermore, for any dark energy model to be viable it is necessary the existence of the matter-dominated and radiation-dominated eras before the dark energy era. In the framework of dynamical systems these matter$/$radiation solutions are described in terms of critical points of the autonomous system that are required to be unstable points in order to allow the transition to the dark energy era. For the dark energy model studied in this paper, we have found both the critical points describing the standard radiation$/$matter eras, as well as new scaling solutions representing the scaling radiation$/$matter epochs. Then, from the stability analysis we have obtained that all them are unstable points (saddle points). Furthermore, since the scaling radiation$/$matter era is characterised by the presence of a small portion of dark energy during the radiation$/$matter era, there could be induced effects on the CMB power spectrum, and also on the matter power spectrum at present through the reduction of the matter fluctuation variance $\sigma_{8}$ because the suppressed growth rate of matter perturbations \cite{Amendola:1999dr,Amendola:1999er,Davari:2019tni}. Therefore, these special features of the present model lead to detectable observational signatures at early and late times that may allow to distinguish it from the $\Lambda$CDM model \cite{Albuquerque:2018ymr,Hill:2020osr}. To further investigate these specific observational signatures using the current CMB and LSS full data set, and to constrain more tightly the free parameters of the model, it is required a detailed analysis of the perturbations around the cosmological background \cite{Amendola:1999er,Davari:2019tni,DiValentino:2019ffd}.
 
We would also like to note that for the present model to be a good candidate for description of our Universe, it must be verified that it is free from any theoretical pathologies, such as ghost, gradient and tachyonic instabilities, through a rigorous stability analysis in the presence of matter fields \cite{DeFelice:2016ucp}, as well as, it is necessary to perform a detailed comparison with all the cosmological observational data, e.g. SNIa, BAO, CMB, LSS, etc \cite{Davari:2019tni,DiValentino:2019ffd}, and Solar System data, after extracting spherically symmetric solutions \cite{Iorio:2012cm}. These necessary studies lie beyond the scope of the present work and thus are left for separated projects \cite{Gonzalez-Espinoza:2021mwr}.




\begin{acknowledgements}
M. Gonzalez-Espinoza acknowledges support from PUCV. G. Otalora acknowldeges DI-VRIEA for financial support through Proyecto Postdoctorado $2020$ VRIEA-PUCV.
\end{acknowledgements}

\bibliographystyle{spphys}       
\bibliography{bio}   

\end{document}